\documentclass[journal]{IEEEtran}
\IEEEoverridecommandlockouts

\usepackage{cite}
\usepackage{url}
\usepackage{balance}
\usepackage{adjustbox}
\usepackage{subfigure}
\usepackage{tabularx} 
\usepackage{pifont}
\newcommand{\cmark}{\ding{51}}%
\usepackage{amsmath}
\usepackage{array}
\usepackage{multirow}
\usepackage{rotating}
\usepackage{xcolor}
\PassOptionsToPackage{table,xcdraw}{xcolor}

\newcolumntype{L}[1]{>{\raggedright\let\newline\\\arraybackslash\hspace{0pt}}m{#1}}
\newcolumntype{C}[1]{>{\centering\let\newline\\\arraybackslash\hspace{0pt}}m{#1}}
\newcolumntype{R}[1]{>{\raggedleft\let\newline\\\arraybackslash\hspace{0pt}}m{#1}}

\usepackage{bbding}
\usepackage{dblfloatfix}
\usepackage{color, colortbl}
\usepackage{graphicx}
\usepackage{amssymb}
\usepackage{pifont}
\usepackage{algorithm}
\usepackage[noend]{algpseudocode}
\usepackage[nolist]{acronym}
\usepackage{fancyvrb}
\usepackage{listings}

\usepackage{amsthm}
\newtheorem{thm}{Theorem}

\newtheorem{lem}[thm]{Lemma}

\newtheorem{defn}[thm]{Definition}

\newcommand{\ie}{{\em i.e., }}
\newcommand{\eg}{{\em e.g., }}
\newcommand{\myverb}{\fontsize{10}{48}\usefont{OT1}{lmtt}{b}{n}\noindent }

\ifCLASSINFOpdf
\else
\fi

\begin{document}
\title{Verifying and Monitoring IoTs Network Behavior using MUD Profiles}

\author{Ayyoob~Hamza,
			 Dinesha~Ranathunga,
			 Hassan~Habibi~Gharakheili,  
			 \\Theophilus~A.~Benson, 
			 Matthew~Roughan,
			 and~Vijay~Sivaraman
	\IEEEcompsocitemizethanks{
		\IEEEcompsocthanksitem A. Hamza, H. Habibi Gharakheili, and V. Sivaraman are with the School	of Electrical Engineering and Telecommunications, University of New South Wales, Sydney, NSW 2052, Australia (e-mails: ayyoobhhamza@student.unsw.edu.au, h.habibi@unsw.edu.au, vijay@unsw.edu.au).
		\IEEEcompsocthanksitem D. Ranathunga and M. Roughan are with the the School of Mathematical Sciences, University of Adelaide, SA, 5005, Australia (e-mails: dinesha.ranathunga@adelaide.edu.au, matthew.roughan@adelaide.edu.au).
		\IEEEcompsocthanksitem T. Benson is with the School	of Computer Science and Engineering, Brown University, Providence, RI 02192, USA (e-mail: tab@cs.brown.edu).
		\IEEEcompsocthanksitem This submission is an extended and improved version of our paper presented at the ACM Workshop on IoT S\&P 2018 \cite{IoTSnP18}.
		}
}
	

\maketitle
\begin{abstract}
	IoT devices are increasingly being implicated in cyber-attacks, raising community concern about the risks they pose to critical infrastructure, corporations, and citizens. In order to reduce this risk, the IETF is pushing IoT vendors to develop formal specifications of the intended purpose of their IoT devices, in the form of a Manufacturer Usage Description (MUD),	so that their network behavior in any operating environment can be locked down and verified rigorously.
	
	This paper aims to assist IoT manufacturers in developing and verifying MUD profiles, while also helping adopters of these devices to ensure they are compatible with their organizational policies and track devices network behavior based on their MUD profile. Our first contribution is to develop a tool that takes the traffic trace of an arbitrary IoT device as input and automatically generates the MUD profile for it. We contribute our tool as open source, apply it to 28 consumer IoT devices, and highlight insights and challenges encountered in the process. Our second contribution is to apply a formal semantic framework that not only validates a given MUD profile for consistency, but also checks its compatibility with a given organizational policy. We apply our framework to representative organizations and selected devices, to demonstrate how MUD can reduce the effort needed for IoT acceptance testing.
	Finally, we show how operators can dynamically identify IoT devices using known MUD profiles and monitor their behavioral changes  on their network. 
\end{abstract}


\IEEEpeerreviewmaketitle

\section{Introduction}\label{sec:intro}


The Internet of Things is considered the next technological mega-trend, with wide reaching effects across the business spectrum
\cite{sachs2014}. By connecting billions of every day devices Ð from smart watches to industrial equipment Ð to the Internet, IoT integrates the physical and cyber worlds, creating a host of opportunities and challenges for businesses
and consumers alike. But, increased interconnectivity also increases the risk of using these devices. 

Many connected IoT devices can be found on search engines such as Shodan \cite{Shodan}, and their vulnerabilities exploited at scale. 
For example, Dyn, a major DNS provider, was subjected to a DDoS attack originating from a large IoT botnet comprising thousands of compromised IP-cameras \cite{Dyn16}. IoT devices, exposing TCP/UDP ports to arbitrary local endpoints within a home or enterprise, and to remote entities on the wider Internet, can be used by inside and outside attackers to reflect/amplify attacks and to infiltrate otherwise secure networks\cite{Wisec17}.
IoT device security is thus a top concern for the Internet ecosystem.

These security concerns have prompted standards bodies to provide guidelines for the Internet community to build secure IoT devices and services \cite{DHS16,NIST16,ENISA17}, and for regulatory bodies (such as the US FCC) to control their use \cite{FCC16}. The focus of our work is an IETF proposal called Manufacturer Usage Description (MUD) \cite{ietfMUD18} which provides the first formal framework for IoT behavior that can be rigorously enforced. This framework requires manufacturers of IoTs to publish a behavioral profile of their device, as they are the ones with best knowledge of how their device will behave when installed in a network; for example, an IP camera may need to use DNS and DHCP on the local network, and communicate with NTP servers and a specific cloud-based controller in the Internet, but nothing else. Such requirements vary across IoTs from different manufacturers. Knowing each device's requirements will allow network operators to impose a tight set of access control list (ACL) restrictions for each IoT device in operation, so as to reduce the potential attack surface on their network.

The MUD proposal hence provides a light-weight model to enforce effective baseline security for IoT devices by allowing a network to auto-configure the required network access for 
the devices, so that they can perform their intended functions without having unrestricted network privileges. 

MUD is a new and emerging paradigm, and there is little collective wisdom today on how manufacturers should develop behavioral profiles of their IoT devices, or how organizations should use these profiles to secure their network and monitor the runtime behaviour of IoT devices. 
Our preliminary work in \cite{hamza2018} was one of the first attempts to address these shortcomings.
This paper\footnote{This project was supported by Google Faculty Research Awards Centre of Excellence for Mathematical and Statistical Frontiers (ACEMS).} significantly expands on our prior work by
proposing an IoT device classification framework which uses observed traffic traces and incrementally compares them with known IoT MUD signatures.
We use this framework and trace data captured over a period of six months from a test-bed comprising
of 28 distinct IoT devices  to identify 
(a) legacy IoT devices without vendor MUD support; 
(b) IoT devices with outdated firmware; and (c) IoT devices which are potentially compromised. To the best of our knowledge, this is the first attempt to automatically generate MUD profiles, formally check their consistency and compatibility with an organizational policy, prior to deployment.
In summary, our contributions are:

\begin{itemize}
	\item{We instrument a tool to assist IoT manufacturers to generate MUD profiles. Our tool takes as input the packet trace containing the operational behavior of an IoT device, and generates as ouput a MUD profile for it. We contribute our tool as open source\cite{mudgenerator}, apply it to 28 consumer IoT devices, and highlight insights and challenges encountered in the process.}
	\item{We apply a formal semantic framework that not only validates a given MUD profile for consistency, but also checks its compatibility with a given organizational policy. We apply our semantic framework to representative organizations and selected devices, and demonstrate how MUD can greatly simplify the process of IoT acceptance into the organization.}
	\item{We propose an IoT device classification framework using observed traffic traces and known MUD signatures to dynamically identify IoT devices and monitor their behavioral changes in a network.}
\end{itemize}

The rest of the paper is organized as follows: \S\ref{sec:prior} describes relevant background work on IoT security and formal policy modeling. \S\ref{sec:mud} describes our open-source tool for automatic MUD profile generation. Our verification framework for MUD policies is described in \S\ref{sec:pdn}, followed by evaluation of results. We describe our 
IoT device classification framework in \S\ref{sec:dynProf} and demonstrate its use to identify and monitor IoT behavioral changes within a network.
We conclude the paper in \S\ref{sec:con}.

\section{Background and Related Work}\label{sec:prior}

Securing IoT devices has played a secondary role to innovation, i.e., creating new IoT functionality (devices and services). This neglection of security has created a substantial safety and economic risks for the Internet \cite{Survey17}. 
Today many manufacturer IoT devices lack even the basic security measures \cite{IoTSnp17}
and network operators have poor visibility into the network activity of their connected devices 
hindering the application of access-control policies to them \cite{CiscoReport18}.
IoT botnets continue to grow in size and sophistication and
attackers are leveraging them to launch 
large-scale DDoS attacks \cite{f5Labs17}; devices such as baby monitors, refrigerators and smart 
plugs have been hacked and controlled remotely \cite{sivaraman2016smart}; and many organizational assets such as cameras are being accessed publicly \cite{SonyCam,Insecam18}. 

Existing IoT security guidelines and recommendations \cite{DHS16,NIST16,ENISA17,FCC16} are largely qualitative and subject to human interpretation, and therefore unsuitable for automated and rigorous application. The IETF MUD specification \cite{ietfMUD18} on the other hand defines a formal framework to capture device run-time behavior, and is therefore amenable to rigorous evaluation. 
IoT devices also often have a small and recognizable pattern of communication (as demonstrated in our previous work \cite{SmartCity17}).
Hence, the MUD proposal allows IoT device behaviour to be captured succinctly,
verified formally for compliance with organizational policy, and assessed at run-time for anomalous behavior that could indicate an ongoing cyber-attack. 

A valid MUD profile contains a root object called ``access-lists'' container \cite{ietfMUD18} which comprise of several access control entries (ACEs), serialized in JSON format. Access-lists are explicit in describing the direction of communication, \ie \textit{from-device} and \textit{to-device}. 
Each ACE matches traffic on source/destination port numbers for TCP/UDP, and type and code for ICMP. 
The MUD specifications also distinguish \textit{local-networks} traffic from \textit{Internet} communications.

\begin{figure}[t]
	\centering
	\includegraphics[scale=0.4]{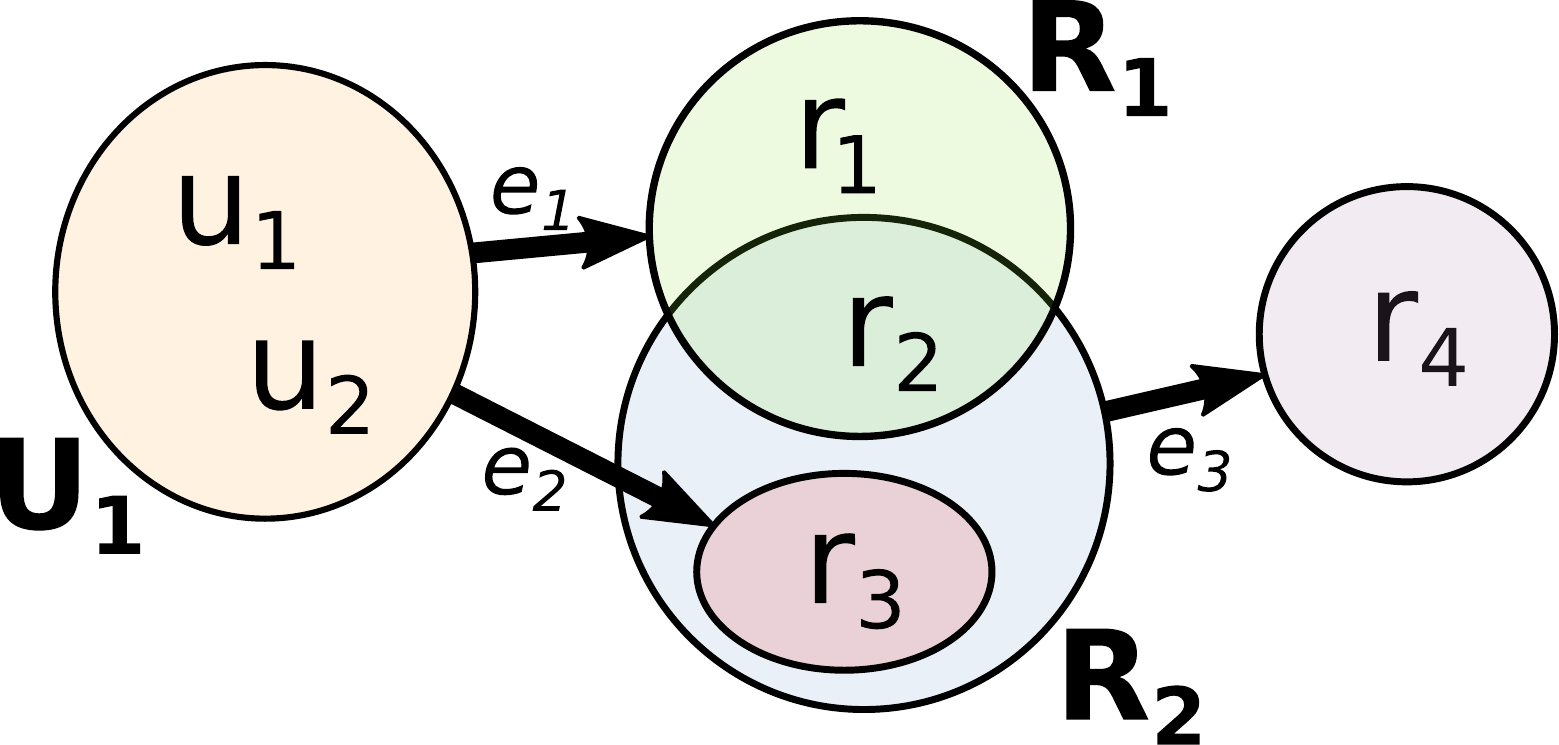}
	\caption{A metagraph consisting of six variables, five sets and three edges.}
	\label{fig:mg1}
	\vspace{-4mm}
\end{figure}

We provide here a brief background on the formal modeling and verification framework used in this paper. We begin by noting that the lack of formal policy modeling in current network systems contribute to frequent misconfigurations \cite{wool2010,ranathunga2016T,ranathunga2016G}. We use the concept of a {\em metagraph}, which is a generalized graph-theoretic structure that offers rigorous formal foundations for modeling and analyzing communication-network policies in general. A metagraph is a directed graph between a collection of sets of ``atomic'' elements \cite{basu2007}. Each set is a node in the graph and each directed edge represents the relationship between two sets. Fig.~\ref{fig:mg1} shows an example where a set of users ($U_1$) are related to sets of network resources ($R_1$, $R_2$, $R_3$) by the edges $e_1, e_2$ and $e_3$ describing which user $u_i$ is allowed to access resource $r_j$.

Metagraphs can also have attributes associated with their edges. An example is a {\em conditional metagraph} which includes propositions -- statements that may be true or false -- assigned to their edges as qualitative attributes \cite{basu2007}. The generating sets of these metagraphs are partitioned into a variable set and a proposition set. A conditional metagraph is formally defined as follows:
\begin{defn}[Conditional Metagraph] 
	A conditional metagraph is a metagraph $S$=$\langle X_p \cup X_v, E \rangle$
	in which $X_p$ is a set of propositions and $X_v$ is a set of variables,
	and:\\
	\indent 1. at least one vertex is not null, \ie $\forall e' \in E, V_{e'} \cup W_{e'} \neq \phi$\\ 
	\indent 2. the invertex and outvertex of each edge must be disjoint, \ie $X = X_v \cup X_p$ with $X_v \cap X_p = \phi$\\
	\indent 3. an outvertex containing propositions cannot contain other elements, \ie 
	$\forall p \in X_p, \forall e' \in E$, if $p \in W_{e'}$, then $W_{e'}={p}$. 
	\label{def:metagraph}
\end{defn}

Conditional metagraphs enable the specification of stateful network-policies and have several useful operators. These operators readily allow one to analyze MUD policy properties like consistency. 

The MUD proposal defines how a MUD profile needs to be fetched. The MUD profile will be downloaded using a MUD url (\eg via DHCP option). For legacy devices already in production networks, MUD specifications suggest to create a mapping of those devices to their MUD url. Therefore, in this paper, we develop a method (in \S\ref{sec:dynProf}) for automatic device identification using MUD profiles to reduce the complexity of manual mapping a device to its corresponding MUD-url.




Past works have employed machine learning to classify IoT devices for asset management \cite{meidan2017, sivanathan2018}.  Method in \cite{meidan2017}  employs over 300 attributes (packet-level and flow-level), though the most influential ones are minimum, median, and average of packet volume, Time-To-Live (TTL), the ratio of total bytes transmitted and received, and the total number of packets with RST flag reset. Work in \cite{sivanathan2018} proposes to use features with less computation cost at runtime. Existing Machine learning based proposals need to re-train their model when a new device type is added -- this limits the usability in terms of not being able to transfer the models across deployments.


While all the above works make important contributions,
they do not leverage the MUD proposal, which the IETF is pushing for vendors to adopt. We overcome the shortfall by developing an IoT device classification framework which dynamically compares the device traffic traces (run-time network behavior) with known static IoT MUD signatures.
Using this framework, we are able to identify (a) legacy IoT devices without vendor MUD support; 
(b) IoT devices with outdated firmware; and (c) IoT devices which are potentially compromised.


\section{MUD Profile Generation}\label{sec:mud}

The IETF MUD specification is still evolving as a draft. Hence, IoT device manufacturers have not yet provided MUD profiles for their devices. 
We, therefore, developed a tool -- \textit{MUDgee} -- which automatically generates a MUD profile for an IoT device from its traffic trace in order to make this process faster, cheaper and more accurate. In this section, we describe the structure of our open source tool \cite{mudgenerator}, apply it to traces of 28 consumer IoT devices, and highlight insights. 

We captured traffic flows for each IoT device during a six month observation period, to generate our MUD rules. The IETF MUD draft allows both `allow' and `drop' rules.  In our work, instead, we generate profiles that follow a
whitelisting model (\ie only `allow' rules with default `drop'). Having a combination of `accept' and `drop' rules requires a notion of rule priority (\ie order) and is not supported by the current IETF MUD draft. For example, Table~\ref{table:bpmrules} shows traffic flows observed for a Blipcare blood pressure monitor. The device only generates traffic whenever it is used. 
It first resolves its intended server at  {\myverb{tech.carematrix.com}} by exchanging a DNS query/response with the default gateway (\ie the top two flows).
It then uploads the measurement to its server operating on TCP port 8777 (described by the bottom two rules).

\begin{table}[t!]
	\small
	\caption{Flows observed for Blipcare BP monitor (*: wildcard, proto: Protocol, sPort: source port number, dPort: destination port number).} 
	\vspace{-2mm}
	\centering 
	\begin{adjustbox}{max width=0.48\textwidth}
		\renewcommand{\arraystretch}{1.3}
		\begin{tabular}{l l l l l} 
			\hline\hline 
			\multicolumn{1}{p{0.7cm}}{\raggedleft \textbf{Source} } & \multicolumn{1}{p{0.7cm}}{\raggedleft \textbf{Destination}} & 
			\multicolumn{1}{p{0.6cm}}{\raggedleft \textbf{proto}} & \multicolumn{1}{p{0.85cm}}{\raggedleft \textbf{sPort} } & 
			\multicolumn{1}{p{0.85cm}}{\raggedleft \textbf{dPort}}\\
			\hline 
			{\myverb{*}} & {\myverb{192.168.1.1}} & {\myverb{17}} & {\myverb{*}} & {\myverb{53}}\\
			{\myverb{192.168.1.1}} & {\myverb{*}} & {\myverb{17}}& {\myverb{53}} & {\myverb{*}} \\
			{\myverb{*}} & {\myverb{tech.carematix.com}} & {\myverb{6}} & {\myverb{*}}  & {\myverb{8777}}\\
			{\myverb{tech.carematix.com}} & {\myverb{*}} & {\myverb{6}} & {\myverb{8777}} & {\myverb{*}}\\
			\hline 
		\end{tabular}
	\end{adjustbox}
	\label{table:bpmrules} 
	\vspace{-2mm}
\end{table}

\subsection{MUDgee Architecture}

{\em MUDgee} implements a programmable virtual switch (vSwitch) with a header inspection engine attached and plays an input PCAP trace (of an arbitrary IoT device) into the switch. {\em MUDgee} has two separate modules; (a) captures and tracks all TCP/UDP flows to/from device, and (b) composes a MUD profile from the flow rules. We describe these two modules in detail below.  

\smallskip
\noindent \textbf{Capture intended flows:} Consumer IoT devices use services provided by remote cloud servers and also expose services to local hosts (\eg a mobile App). We track (intended) both remote and local device communications using separate flow rules to meet the
MUD specification requirements.

\begin{figure}[t]
	\centering
	\includegraphics[scale=0.3]{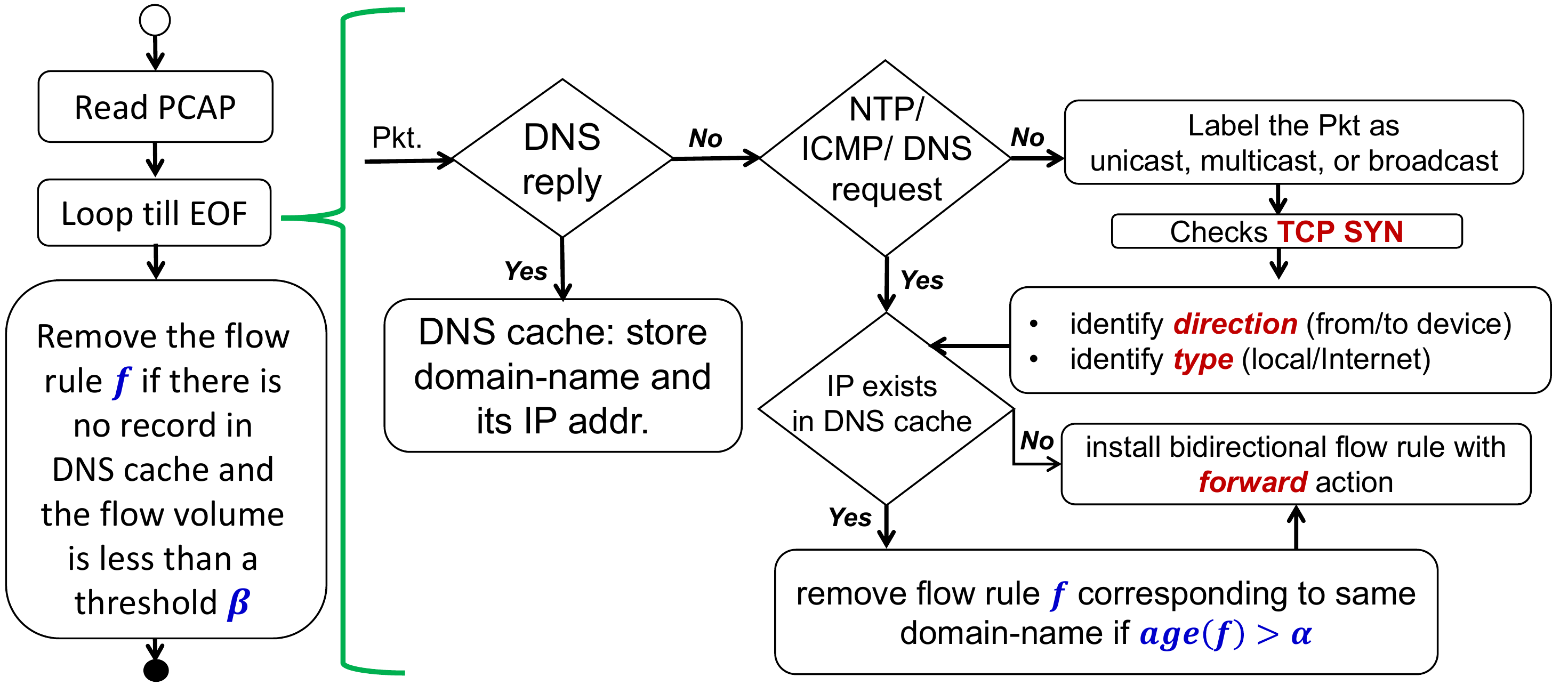}
	\caption{Algorithm for capturing device flows and inserting reactive rules.}
	\label{fig:flowidentifier}
	\vspace{-0.35cm}
\end{figure}

It is challenging to capture services (\ie especially those operating on non-standard TCP/UDP ports) that a device is either accessing or exposing. This is because local/remote services operate on static port numbers whereas source port numbers are dynamic (and chosen randomly) for different flows of the same service. We note that it is trivial to deduce the service for TCP flows by inspecting the SYN flag, but not so easy for UDP flows. We, therefore, developed an algorithm (Fig.~\ref{fig:flowidentifier}) to capture bidirectional flows for an IoT device. 

We first configure the vSwitch with a set of proactive rules, each with a specific action (\ie``forward'' or ``mirror'') and a priority (detailed rules can be found in our technical report \cite{hamza2018}). 
Proactive rules with a `mirror' action will feed the header inspection engine with a copy of the matched packets. Our inspection algorithm, shown in Fig.~\ref{fig:flowidentifier}, will insert a corresponding reactive rule into the vSwitch. 

Our algorithm matches a DNS reply to a top priority flow
and extracts and stores the 
domain name and its  associated IP address in a DNS cache. 
This cache is dynamically updated upon arrival of a DNS reply matching an existing request.

\begin{figure*}[!t]
	\vspace{-4mm}
	\begin{center}
		\mbox{
			\subfigure[TP-Link camera.]{
				{\includegraphics[width=0.47\textwidth,height=0.19\textheight]{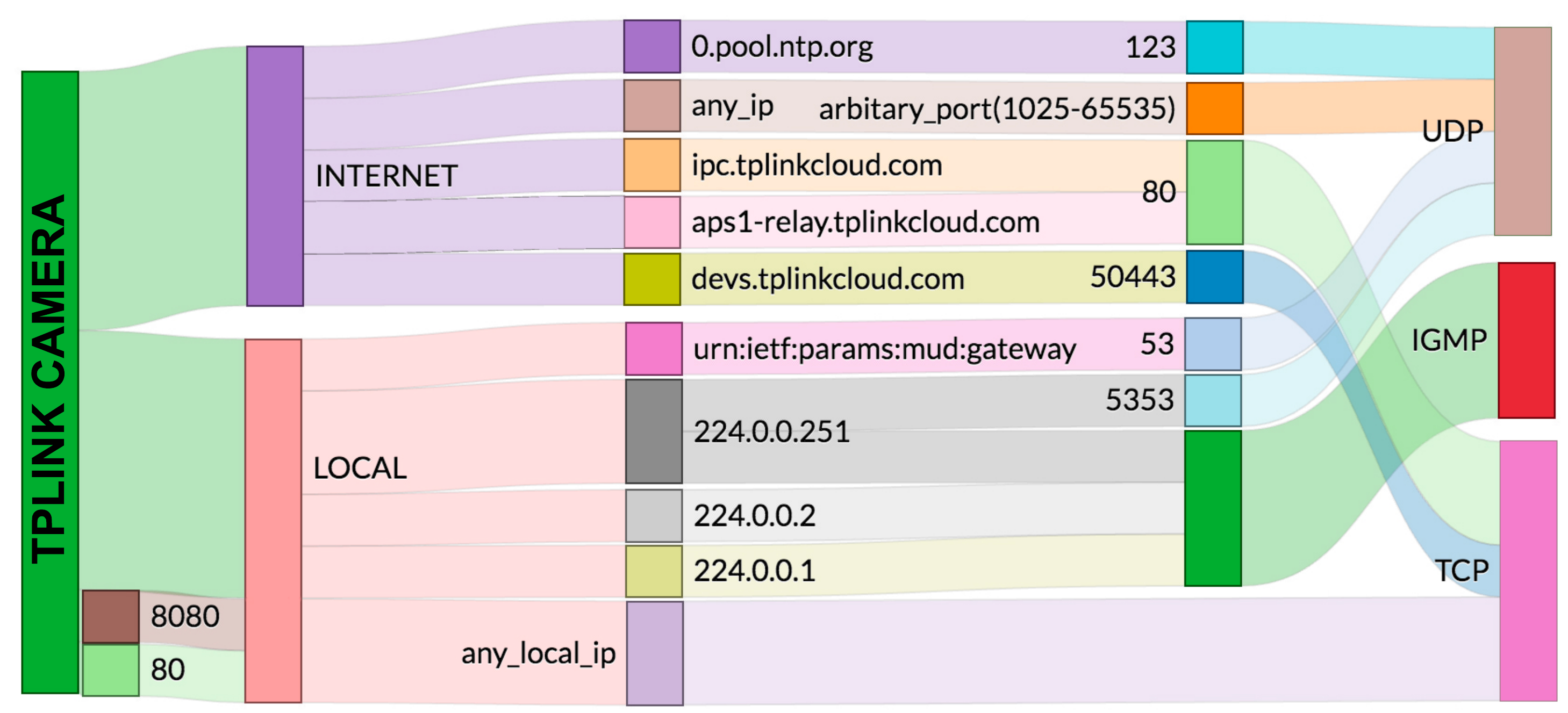}}\quad
				\label{fig:sankeyTPlinkCam}
			}
			\hspace{-4mm}
			\subfigure[Amazon Echo (see Listing 1 for description of domain\_set1-3).]{
				{\includegraphics[width=0.47\textwidth,height=0.19\textheight]{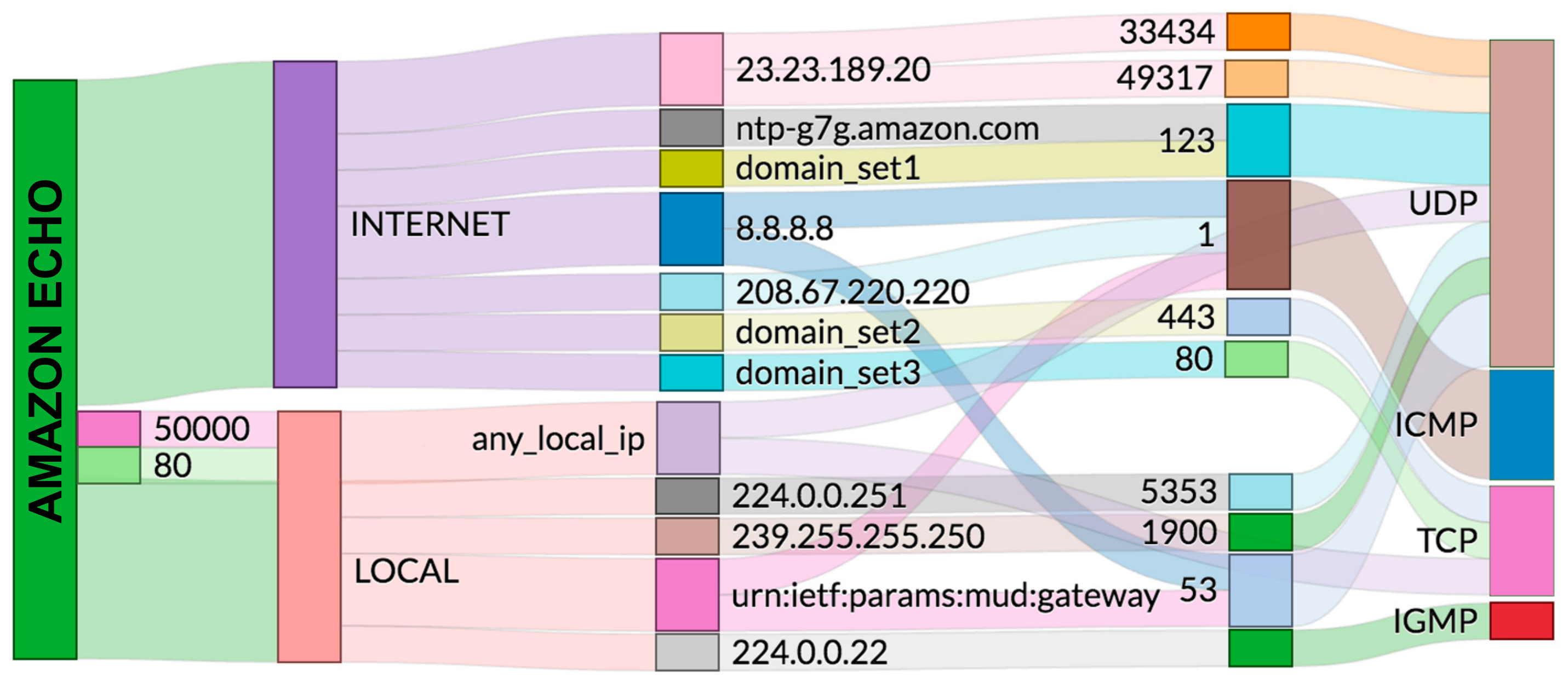}}\quad
				\label{fig:sankeyEcho}
			}
		}
		\vspace{-4mm}
		\caption{Sankey diagrams of MUD profiles for: (a) TP-Link camera, and (b) Amazon Echo. }
		\vspace{-1mm}
		\label{fig:Snakey}
	\end{center} 
	\vspace{-4mm}
\end{figure*}

The MUD specification also requires the segregation of traffic to and from a device for both local and Internet communications.
Hence, our algorithm assigns a unique priority to the reactive rules associated with each of the groups: from-local, to-local, from-Internet and to-Internet.
We use a specific priority for flows that contain a TCP SYN to identify if the device or the remote entity initiated the communication.

\smallskip
\noindent \textbf{Flow translation to MUD:}
{\em MUDgee} uses the captured traffic flows to generate a MUD profile for each device. We convert each flow to a MUD ACE by considering the following:

\smallskip
\textit{\textbf{Consideration 1}}: We reverse lookup the IP address of the remote endpoint and identify the associated domain name (if any), using the DNS cache.

\textit{\textbf{Consideration 2}}: Some consumer IoTs, especially IP cameras, typically use the \ac{STUN} protocol to verify that the user's mobile app can stream video directly from the camera over the Internet. If a device uses the STUN protocol over UDP, we must allow all UDP traffic to/from Internet servers because the STUN servers often require the client device to connect to different IP addresses or port numbers. 

\textit{\textbf{Consideration 3}}: We observed that several smart IP cameras communicate with many remote servers operating on the same port (\eg Belkin Wemo switch). However, no DNS responses were found corresponding to the server IP addresses. So, the device must obtain the IP address of its servers via a non-standard channel (\eg the current server may instruct the device with the IP address of the subsequent server).  If a device communicates with several remote IP addresses (\ie more than our threshold value of five), all operating on the same port, we allow remote traffic to/from any IP addresses (\ie *) on that specific port number.

\textit{\textbf{Consideration 4}}: Some devices (\eg TPLink plug) use the default gateway as the DNS resolver, and others (\eg Belkin WeMo motion) continuously ping the default gateway. 
The existing MUD draft maps local communication to fixed IP addresses through the controller construct.
We consider the local gateway to act as the controller, and use the name-space {\myverb{urn:ietf:params:mud:gateway}} for the gateway.

\smallskip
The generated MUD profiles of the 28 consumer IoT devices in our test bed are listed in Table~\ref{table:IoTdevices} and are publicly available at: 
{\myverb{https://iotanalytics.unsw.edu.au/mud/}}.

\vspace{-2mm}
\subsection{Insights and challenges} 
The Blipcare BP monitor is an example device with static functionalities. 
It exchanges DNS queries/responses with the local gateway 
and communicates with a single domain name over TCP port 8777. 
So its behavior can be locked down to a limited set of static flow rules. 
The majority of IoT devices that we tested (\ie 22 out of 28) fall into this category (marked in green in Table~\ref{table:IoTdevices}).   

We use  Sankey diagrams (shown in Fig.~\ref{fig:Snakey}) to represent the MUD profiles in a human-friendly way.
The second category of our generated MUD profiles is 
exemplified by Fig.~\ref{fig:sankeyTPlinkCam}.
This Sankey diagram shows how
the TP-Link camera accesses/exposes limited ports on the local network.
The camera gets its DNS queries resolved, discovers local network using mDNS over UDP 5353, probes members of certain multicast groups using IGMP, and exposes two TCP ports 80 (management console) and 8080 (unicast video streaming) to local devices. 
All these activities can be defined by a tight set of ACLs. 

But, over the Internet, the camera communicates to its STUN server,
accessing an arbitrary range of IP addresses and port numbers shown by the top flow.
Due to this communication, the functionality of this device can only be loosely defined.
Devices that fall in to this category (\ie due to the use of STUN protocol),
are marked in blue in Table~\ref{table:IoTdevices}.
The functionality of these devices can be more tightly defined
if manufacturers of these devices
configure their STUN servers to operate on a 
specific set of endpoints and port numbers, instead
of a broad and arbitrary range.

Amazon Echo, represents devices with complex 
and dynamic functionality, 
augmentable using custom recipes or skills. 
Such devices (marked in red in Table~\ref{table:IoTdevices}), can communicate
with a growing range of endpoints on the Internet, 
which the original manufacturer cannot define in advance. For
example, our Amazon Echo interacts with the Hue lightbulb

\begin{table}[t!]
	\small
	\caption{List of IoT devices for which we have generated MUD profiles. Devices with purely static functionality are marked in green. Devices with static functionality that is loosely defined (\eg due to use of STUN protocol) are marked in blue. Devices with complex and dynamic functionality are marked in red.} 
	\centering 
	\begin{adjustbox}{max width=0.495\textwidth}
		\begin{tabular}{|L{2cm}|L{6cm}|}
			\hline 
			\multicolumn{1}{|c|} {\textbf{Type}} & 	\multicolumn{1}{|c|} {\textbf{IoT device}}\tabularnewline
			\hline 
			Camera & {\color{green!70!blue}Netatmo Welcome, Dropcam, Withings Smart Baby Monitor, Canary camera}, {\color{blue}TP-Link Day Night Cloud camera, August doorbell camera, Samsung SmartCam, Ring doorbell, Belkin NetCam}  \tabularnewline
			\hline 
			Air quality sensors & {\color{green!70!blue}Awair air quality monitor, Nest smoke sensor, Netatmo weather station}  \tabularnewline
			\hline 	
			Healthcare devices & {\color{green!70!blue}Withings Smart scale, Blipcare Blood Pressure meter, Withings Aura smart sleep sensor}  \tabularnewline
			\hline 			
			Switches and Triggers & {\color{green!70!blue}iHome power plug, WeMo power switch, TPLink plug, Wemo Motion Sensor} \tabularnewline
			\hline 		
			Lightbulbs & {\color{green!70!blue}Philips Hue lightbulb, LiFX bulb} \tabularnewline
			\hline 				
			Hub & {\color{red}Amazon Echo}, {\color{green!70!blue}SmartThings}\tabularnewline
			\hline
			Multimedia& {\color{green!70!blue}Chromecast}, {\color{red}Triby Speaker}\tabularnewline
			\hline
			Other & {\color{green!70!blue}HP printer, Pixstar Photoframe, Hello Barbie} \tabularnewline
			\hline 		
		\end{tabular}
	\end{adjustbox}
	\vspace{-6mm}
	\label{table:IoTdevices} 
\end{table}

\noindent in our test bed by communicating with
{\myverb{meethue.com}} over TCP 443.
It also contacts the news website {\myverb{abc.net.au}} when prompted by the user. 
For these type of devices, the biggest challenge is how manufacturers can dynamically update their MUD profiles to match the device capabilities. 
But, even the initial MUD profile itself can help setup
a minimum network-communication permissions set
that can be amended over time.

\begin{figure}[t!] 
	\begin{lstlisting}[caption={Example list of domains accessed by Amazon Echo corresponding to Figure 2(b).}, label=fig:domains, breaklines=true, basicstyle=\small\fontfamily{cmtt}\selectfont]
	domain_set1: 
	0.north-america.pool.ntp.org, 
	1.north-america.pool.ntp.org, 
	3.north-america.pool.ntp.org
	domain_set2: 
	det-ta-g7g.amazon.com, 
	dcape-na.amazon.com, 
	softwareupdates.amazon.com,
	domain_set3: 
	kindle-time.amazon.com, 
	spectrum.s3.amazonaws.com, 
	d28julafmv4ekl.cloudfront.net, 
	live-radio01.mediahubaustralia.com, 
	amzdigitaldownloads.edgesuite.net, 
	www.example.com
	\end{lstlisting}
	\vspace{-7mm}
\end{figure} 

\section{MUD profile verification}
\label{sec:pdn}

Network operators should not allow a device to be installed in their network,
without first checking its compatibility with the organisation's security policy.
We've developed a tool -- {\em MUDdy} -- which can help with the task.
{\em MUDdy} can check an IoT device's MUD profile is correct syntactically and semantically
and ensure that only devices which are compliant and have MUD signatures that adhere to the IETF proposal
are deployed in a network.

\vspace{-3mm}
\subsection{Syntactic correctness}

A MUD profile comprises of a YANG model that describes device-specific network behavior.
In the current version of MUD, this model is serialized using JSON \cite{ietfMUD18}
and this serialisation is limited to a few YANG modules (\eg  ietf-access-control-list).
{\em MUDdy} raises an invalid syntax exception when parsing a MUD profile if it detects 
any schema beyond these permitted YANG modules.

{\em MUDdy} also rejects MUD profiles containing IP addresses with local significance.
The IETF advises MUD-profile publishers to utilise the high-level abstractions provided in the MUD proposal
and avoid using hardcoded private IP addresses \cite{ietfMUD18}.
{\em MUDdy} also discards MUD profiles containing access-control actions other than `accept' or `drop'.
\vspace{-3mm}

\subsection{Semantic correctness}

Checking a MUD policy's syntax partly verifies its correctness. 
A policy must additionally be semantically correct; 
so we must check a policy, for instance, for inconsistencies.

Policy inconsistencies can produce unintended consequences \cite{wool2004}
and in a MUD policy, inconsistencies can stem from 
(a) overlapping rules with different access-control actions; and/or 
(b) overlapping rules with identical actions. 
The MUD proposal excludes rule ordering,
so, the former describes ambiguous policy-author intent ({\ie intent-ambiguous rules}).
In comparison, the latter associates a clear (single) outcome and
describes {\em redundancies}.
Our adoption of an application-whitelisting model prevents the former by design, but, 
redundancies are still possible and need to be checked.

{\em MUDdy} models a MUD policy using a metagraph underneath.
This representation enables us to use Metagraph algebras \cite{basu2007} to precisely check the policy model's consistency. 
It's worth noting here that past works \cite{al2005}  classify policy
consistency based on the level of policy-rule overlap. But,
these classifications are only meaningful when the policy-rule
order is important (\eg in a vendor-device implementation).
However, rule order is not considered in the IETF MUD proposal
and it is also generally inapplicable in the 
context of a policy metagraph.
Below is a summary description of the process we use to check the consistency of a policy model.

\subsubsection{Policy modeling}\label{sec:modeling}

\begin{figure}[t]
	\centering
	\includegraphics[scale=0.20]{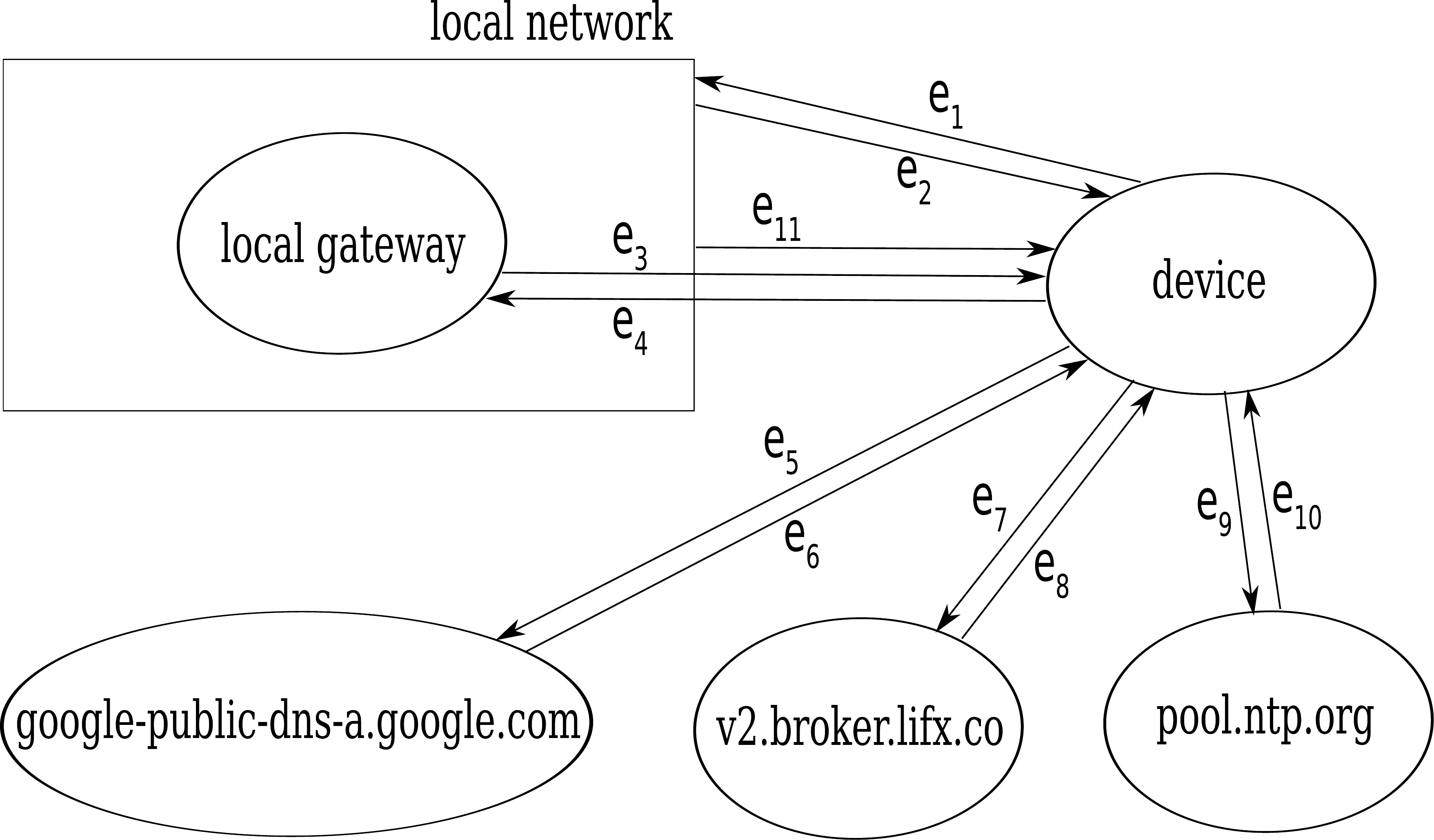}
	\caption{Metagraph model of a LiFX bulb's MUD policy. The policy describes permitted traffic flow behavior. Each edge label has attached a set of propositions of the metagraph. For example $e_4$=$\{protocol=17,UDP.dport=53,UDP.sport=0-65535,action=accept \}$.}
	\label{fig:cmg_bulb}
	\vspace{-5mm}
\end{figure}

Access-control policies are often represented using the five-tuple: source/destination address, protocol, source/destination ports \cite{cisco2013,juniper2016,palo2017}. We construct MUD policy metagraph models leveraging this idea. Fig.~\ref{fig:cmg_bulb} shows an example for a Lifx bulb. Here, the source/destination addresses are represented by the labels {\myverb{device}}, {\myverb{local-network}}, {\myverb{local-gateway}} and a domain-name (\eg {\myverb{pool.ntp.org}}). Protocol and ports are propositions of the metagraph.

\subsubsection{Policy definition and verification}\label{sec:definition}

We wrote {\em MGtoolkit} \cite{ranathunga2017} -- a package for implementing metagraphs -- 
to instantiate our policy models. {\em MGtoolkit} is implemented in Python 2.7. 
The API allows users to create metagraphs, apply metagraph operations and evaluate results.

{\em Mgtoolkit} provides a
{\myverb{ConditionalMetagraph}} class which extends a {\myverb{Metagraph}} and 
supports propositions.
The class inherits the members of a {\myverb{Metagraph}}
and additionally supports methods to check consistency.
We use this class to instantiate our MUD policy models and check their consistency.

Our verification of metagraph consistency uses {\em dominance} \cite{basu2007} which
can be introduced constructively as follows: 

\smallskip
\begin{defn}[Edge-dominant Metapath]
	Given a metagraph $S$=$ \langle X, E \rangle$ for any two sets of elements
	$B$ and $C$ in $X$, a metapath $M(B,C)$ is said to be edge-dominant if no proper
	subset of $M(B,C)$ is also a metapath from $B$ to $C$.
	\label{def:metapath1}
\end{defn}

\begin{defn}[Input-dominant Metapath]
	Given a metagraph $S$=$ \langle X, E \rangle$ for
	any two sets of elements $B$ and $C$ in $X$, a metapath $M(B,C)$ is said to be
	input-dominant if there is no metapath $M'(B',C)$ such that $B' \subset B$.
	\label{def:metapath2}
\end{defn}

In other words, edge-dominance (input-dominance) ensures that none of the
edges (elements) in the metapath are redundant. 
These concepts allow us to define a dominant metapath as per below.
A non-dominant metapath indicates redundancy in the policy
represented by the metagraph. 


\begin{defn}[Dominant Metapath]
	Given a metagraph $S$=$ \langle X, E \rangle$ for any two sets of elements
	$B$ and $C$ in $X$, a metapath $M(B,C)$ is said to be dominant if it is both edge dominant
	and input-dominant.
	\label{def:metapath3}
\end{defn}

\subsubsection{Compatibility with best practices}
\label{sec:bp_check}

MUD policy consistency checks partly verify if it is semantically correct.
In addition, a MUD policy may need to be verified against a local security policy
or industry recommended practices (such as the ANSI/ISA- 62443-1-1), for compliance.
Doing so, is critical when installing an IoT device in a mission-critical network 
such as a SCADA network, where highly restrictive cyber-security practices
are required to safeguard people from serious injury or even death!

We built an example organisational security policy based on SCADA best practice guidelines
to check MUD policy compliance. We chose these best practices because
they offer a wide spectrum of policies representative of various organisations.
For instance, they include policies for the 
highly protected SCADA zone (which, for instance, might run a power plant)
as well as the more moderately-restrictive Enterprise zone.

We define a MUD policy rule to be SCADA (or Enterprise) zone compatible if its 
corresponding traffic flow complies with SCADA (or Enterprise) best practice policy.
For instance, a MUD rule which permits a device to communicate with the local network using DNS
complies with the Enterprise zone policy. But, a rule enabling
device communication with an Internet server using HTTP violates the SCADA zone policy.

Our past work has investigated the problem of policy comparison using formal semantics, 
in the SCADA domain for firewall access-control policies \cite{ranathunga2016P}. 
We adapt the methods and algebras developed there, to also check MUD policies 
against SCADA best practices. Key steps enabling these formal comparisons are summarized below.

Policies are mapped into a unique canonical decomposition.
Policy canonicalisation can be represented through a mapping $c: \Phi
\rightarrow \Theta$, where $\Phi$ is the policy space and $\Theta$ is the canonical space of
policies. All equivalent policies of $\Phi$ map to a
singleton. For $p^X, p^Y \in \Phi$, we note the following (the proof 
follows the definition)
\begin{lem}
	Policies $p^X \equiv p^Y$  iff $c(p^X)=c(p^Y)$.
\end{lem}

MUD policy compliance can be checked by comparing canonical policy components.  For instance

\smallskip
\smallskip
\indent \indent Is $c ( p^{device \to controller} ) = c ( p^{SCADA \to Enterprise} )$ ? 

\smallskip
A notation also useful in policy comparison is that 
policy $P^A$ {\em includes} policy $P^B$. 
In SCADA networks, the notation helps evaluate whether a MUD policy is
compliant with industry-recommended practices in \cite{stouffer2008,byres2005}. 
A violation increases the vulnerability of a SCADA zone to cyber attacks.

We indicate that a policy {\em complies} with another if it is more restrictive or included in and define the following
\begin{defn}[Inclusion]
	A policy $p^X$ is {\em included} in $p^Y$ on $A$ iff $p^X(s)
	\in \{p^Y(s), \phi\}$, \ie $X$ either has the same effect as $Y$ on
	$s$, or denies $s$, for all $s \in A$. We denote inclusion by
	$p^X \subset p^Y$.
	\label{def:includes}
\end{defn}

A MUD policy ($MP$) can be checked against a SCADA best practice policy ($RP$) for compliance using inclusion

\smallskip
\indent \indent \indent \indent \indent  Is $p^{MP} \subset p^{RP}$ ?

\smallskip
The approach can also be used to 
check if a MUD policy complies with an organisation's local security policy, 
to ensure that IoT devices are plug and play enabled, only
in the compatible zones of the network.

\vspace{-3mm}
\subsection{Verification results}
\label{sec:eval}
\begin{table*}[t]
	\caption{MUD policy analysis summary for our test bed IoT devices using {\em Muddy} ( {\em Safe to install?} indicates where in a network (\eg Enterprise Zone, SCADA Zone, DMZ) the device can be installed without violating best practices, DMZ - Demilitarized Zone, Corp Zone - Enteprise Zone). {\em Muddy} ran on a standard desktop computer; \eg Intel Core CPU 2.7-GHz computer with 8GB of RAM running Mac OS X)} 
	\centering 
	\begin{adjustbox}{max width=0.8650\textwidth} 
		\begin{tabular}{lcccccccc} 
			\hline\hline 
			\multicolumn{1}{p{3cm}}{\raggedright Device name} & 
			\multicolumn{1}{p{1.5cm}}{\centering \#MUD profile \\ rules} & 
			\multicolumn{1}{p{1cm}}{\centering \#Redundant \\ rules } & 
			\multicolumn{1}{p{2cm}}{\centering Redundancy checking\\ CPU time (s) } & 
			\multicolumn{1}{p{2cm}}{\centering Compliance checking\\ CPU time (s)} &			
			\multicolumn{1}{p{2cm}}{\centering Safe to \\ install ?} &
			\multicolumn{1}{p{2cm}}{\centering \% Rules violating \\ SCADA Zone} &
			\multicolumn{1}{p{1.5cm}}{\centering \% Rules violating \\ Corp Zone} \\ [0.5ex] 			
			
			\hline
			Blipcare bp & 6  & 0 & 0.06 & 38 & DMZ, Corp Zone & 50 & 0 \\ 
			Netatmo weather & 6 & 0 & 0.04 & 36 & DMZ, Corp Zone & 50 & 0\\ 
			SmartThings hub & 10  & 0 & 1 & 39 & DMZ, Corp Zone & 60 & 0 \\ 
			Hello barbie doll & 12  & 0 & 0.6 & 38 & DMZ, Corp Zone & 33 & 0\\ 
			Withings scale & 15 & 4 & 0.5 & 40 & DMZ, Corp Zone & 33 & 0\\ 
			Lifx bulb & 15 & 0 & 0.8 & 42 & DMZ, Corp Zone & 60 & 0\\ 
			Ring door bell & 16  & 0 & 1 & 39 & DMZ, Corp Zone & 38 & 0\\ 
			Awair air monitor & 16 & 0 & 0.3 & 101 & DMZ, Corp Zone & 50 & 0\\ 
			Withings baby & 18 & 0 & 0.2 & 41 & DMZ, Corp Zone & 28 & 0\\
			iHome power plug & 17 & 0 & 0.1 & 42 & DMZ & 41 & 6\\ 
			TPlink camera & 22 & 0 & 0.4 & 40 & DMZ & 50 & 4\\ 
			TPlink plug & 25 & 0 & 0.6 & 173 & DMZ & 24 & 4\\ 
			Canary camera & 26 & 0 & 0.4 & 61 & DMZ & 27 & 4\\ 
			Withings sensor & 28  & 0 & 0.2 & 71 & DMZ & 29 & 4\\ 
			Drop camera & 28 & 0 & 0.3 & 214 & DMZ & 43 & 11\\ 
			Nest smoke sensor & 32 & 0 & 0.3 & 81 & DMZ & 25 & 3\\
			Hue bulb & 33 & 0 & 2 & 195 & DMZ & 27 & 3\\ 
			Wemo motion & 35  & 0 & 0.4 & 47 & DMZ & 54 & 8\\ 
			Triby speaker & 38 & 0 & 1.5 & 187 & DMZ & 29 & 3\\ 
			Netatmo camera & 40 & 1 & 0.9 & 36 & DMZ & 28 & 2\\ 
			Belkin camera & 46 & 3 & 0.9 & 55 & DMZ & 52 & 11\\ 
			Pixstar photo frame & 46 & 0 & 0.9 & 43 & DMZ & 48 & 28\\ 
			August door camera & 55 & 9 & 0.8 & 38 & DMZ & 42 & 13\\ 
			Samsung camera & 62 & 0 & 1.7 & 193 & DMZ & 39 & 19\\ 
			Amazon echo & 66 & 4 & 3.2 & 174 & DMZ & 29 & 2\\ 
			HP printer & 67  & 10 & 1.8 & 87 & DMZ & 25 & 9\\ 
			Wemo switch & 98 & 3 & 3.1 &  205 & DMZ & 24 & 6\\ 
			Chrome cast & 150 & 24 & 1.1 & 56 & DMZ & 11 & 2\\ 
			\hline
		\end{tabular}
	\end{adjustbox}
	\vspace{-0.4cm}
	\label{table:suc-summary} 
\end{table*}

We ran {\em MUDgee} on a standard laptop computer (\eg Intel Core CPU 3.1 GHz computer with 16GB of RAM running Mac OS X) and generated MUD profiles for 28 consumer IoT devices installed in our test bed.
{\em MUDgee} generated these profiles by parsing a 2.75 Gb PCAP file (containing 4.5 months of packet trace data from our test bed),
within 8.5 minutes averaged per device.
Table~\ref{table:suc-summary} shows a high-level summary of these MUD profiles. 

It should be noted that a MUD profile generated from a device's traffic trace can be incorrect if the device is compromised, as the trace might include malicious flows. 
In addition, the generated MUD profile is limited to the input trace. Our tool can be extended by an API that allows manufacturers to add rules that are not captured in the PCAP trace.

Zigbee, Z-wave and bluetooth technologies are also increasingly being used by IoT devices. Thus, such devices come with a hub capable of communicating with the Internet. In such cases, a MUD profile can be generated only for the hub.

We then ran {\em MUDdy} on a standard desktop computer (\eg Intel Core CPU 2.7-GHz computer with 8GB of RAM running Mac OS X)
to automatically parse the generated MUD profiles and identify inconsistencies within them.
Our adoption of an application whitelisting model restricts inconsistencies to redundancies.
We determined non-dominant metapaths (as per Definition 4) in each policy metagraph
built by {\em MUDdy}, to detect redundancies.
The average times (in milliseconds) taken to find these redundancies
are shown in Table~\ref{table:suc-summary}.

As the table shows, there were for instance,
three redundant rules present in the Belkin camera's MUD policy.
These rules enabled ICMP traffic to the device from the local network as well as 
the local controller, making the policy inefficient. 

Table~\ref{table:suc-summary} also illustrates the results from
our MUD policy best-practice compliance checks.
For instance, a Blipcare blood pressure monitor can be safely installed 
in the Demilitarized zone (DMZ) or the Enterprise zone 
but not in a SCADA zone:
50\% of its MUD rules violate the best practices, 
exposing the zone to potential cyber-attacks.
Policy rules enabling the device to communicate with the Internet directly,
trigger these violations. 

In comparison, an Amazon echo speaker can only be safely installed in a DMZ.
Table~\ref{table:suc-summary} shows that $29$\% of the device's MUD rules violate the best practices if 
it's installed in the SCADA zone. Only $2$\% of the rules violate if it's installed in the Enterprise zone. 
The former violation stems from rules which for instance, enable
HTTP to the device.
The latter is due to rules enabling ICMP to the device from the Internet. 

{\em MUDdy}'s ability to pinpoint to MUD rules which fail compliance,
helps us to identify possible workarounds to overcome the failures.
For instance, in the Belkin camera,
local DNS servers and Web servers can be employed to
localize the device's DNS and Web communications to achieve compliance in the SCADA zone.

\vspace{-3mm}
\subsection{MUD recommendations}

At present, the MUD specification allows both
accept and drop rules but does not specify priority,
allowing ambiguity.
This ambiguity is removed if only accept rules (\ie whitelisting) is used.
Whitelisting means metagraph edges describe enabled traffic flows.
So, the absence of an edge implies two metagraph nodes don't communicate with one another.
But when drop rules are introduced, an edge also describes prohibited traffic flows,
hindering easy visualization and understanding of the policy.
We recommend the MUD proposal be revised to only support explicit `accept' rules. 

The MUD proposal also does not 
support private IP addresses, instead profiles are made readily
transferrable between networks via support for high-level abstractions.
For instance, to communicate with other IoT devices in the network,
abstractions such as {\em same-manufacturer} is provided.

The MUD proposal however, permits the use of public IP addresses.
This relaxation of the rule allows close coupling of policy with network implementation, increasing its sensitivity to network changes.
A MUD policy describes IoT device behavior and should only change when 
its actual behavior alters and not when network implementation changes!
Hardcoded public IP addresses can also lead to accidental DoS of target hosts.
A good example is the DoS of NTP servers at the University of Wisconsin due to hardcoded 
IP addresses in Netgear routers \cite{plonka2003}.
We recommend that support for explicit public IP addresses
be dropped from the MUD proposal. 

\vspace{-2mm}

\section{Checking Run-Time Profile of IoT Devices}\label{sec:dynProf}
In this section, we track the runtime network behavior of IoT devices and map them to a known MUD profile. 
This is needed to manage legacy IoTs which lack vendor support for the MUD standard. To do so, we generate and update a device's runtime behavioral profile (in the form of a tree), and check its ``\textit{similarity}'' to known static MUD profiles provided by manufacturers. We note that computing similarity between two profiles is a non-trivial task.

\begin{figure*}[!t]
	\vspace{-4mm}
	\begin{center}
		\mbox{
			\subfigure[30-minutes of traffic capture.]{
				{\includegraphics[width=0.45\textwidth]{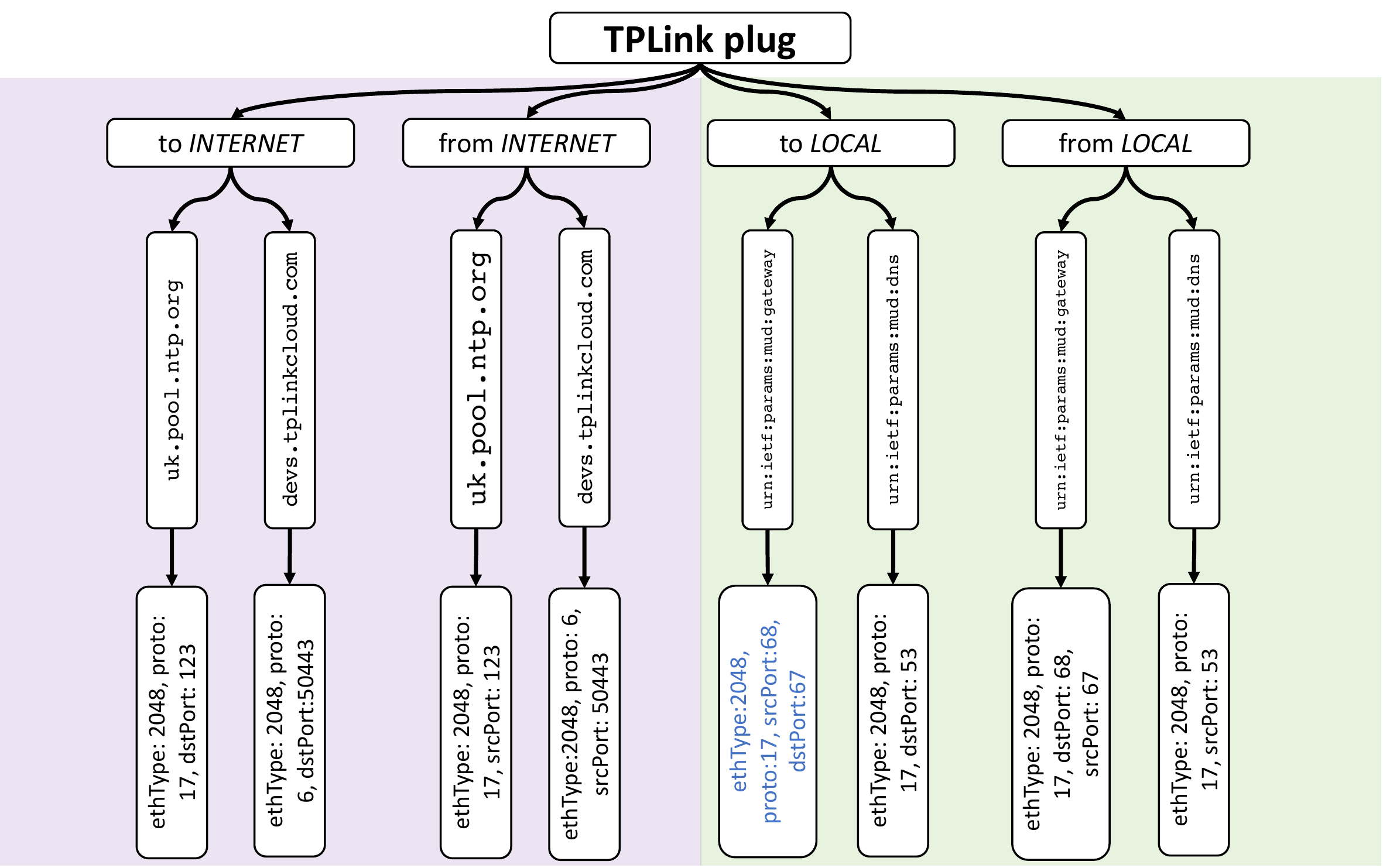}}\quad
				\label{fig:tplinkTree30min}
			}
			\hspace{-4mm}
			\subfigure[480-minutes of traffic capture.]{
				{\includegraphics[width=0.45\textwidth]{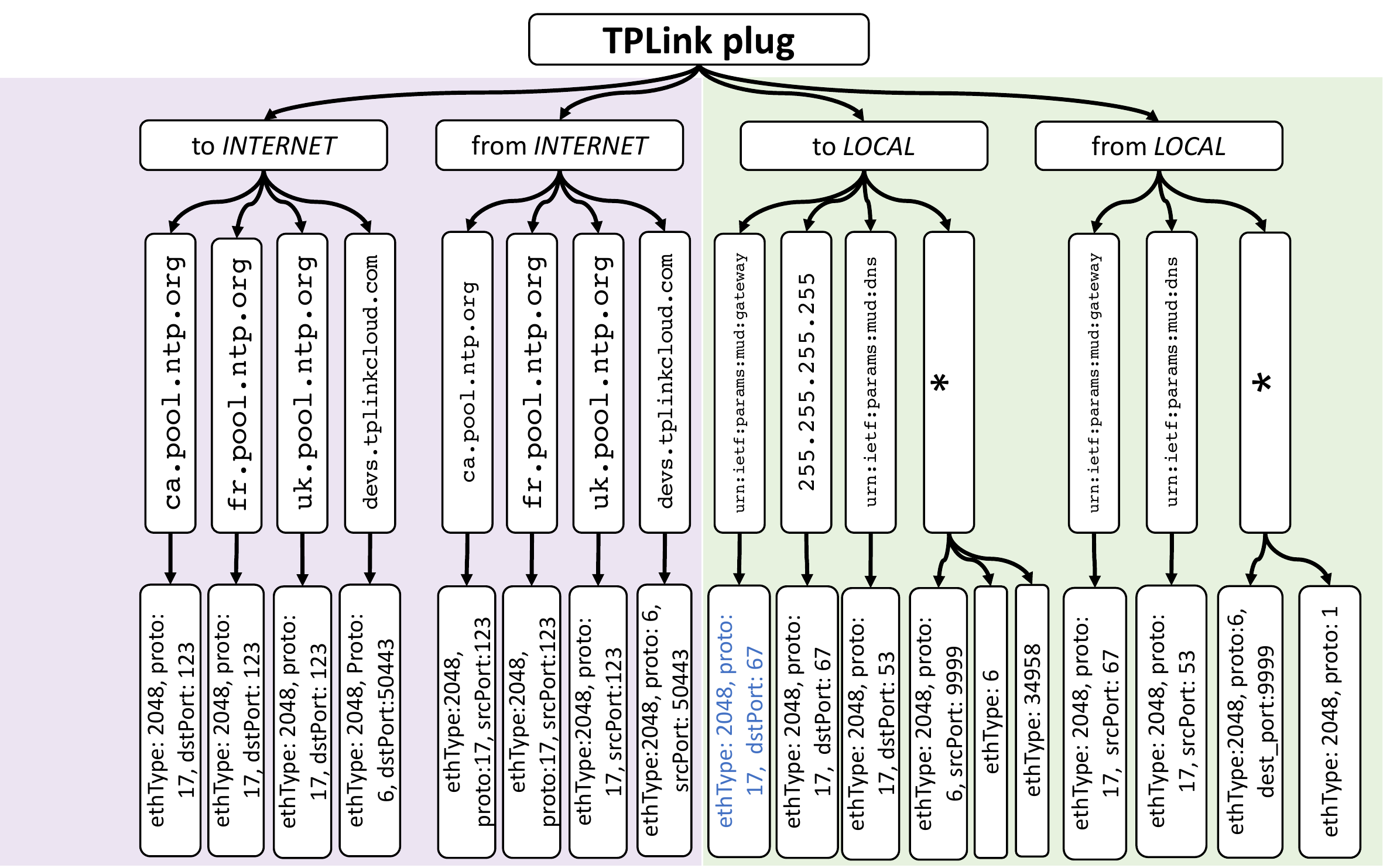}}\quad
				\label{fig:tplinkTree480min}
			}
		}
		\vspace{-4mm}
		\caption{Run-time profile of a TPLink power plug generated at two snapshots in time: (i) after 30 minutes of traffic capture; and (ii) after 8 hours of traffic capture. As observable the profile grows over time by accumulating nodes and edges.}
		\vspace{-1mm}
		\label{fig:tplinkTree}
	\end{center}
\end{figure*}

\begin{figure}[t]
	\centering
	\includegraphics[scale=0.3]{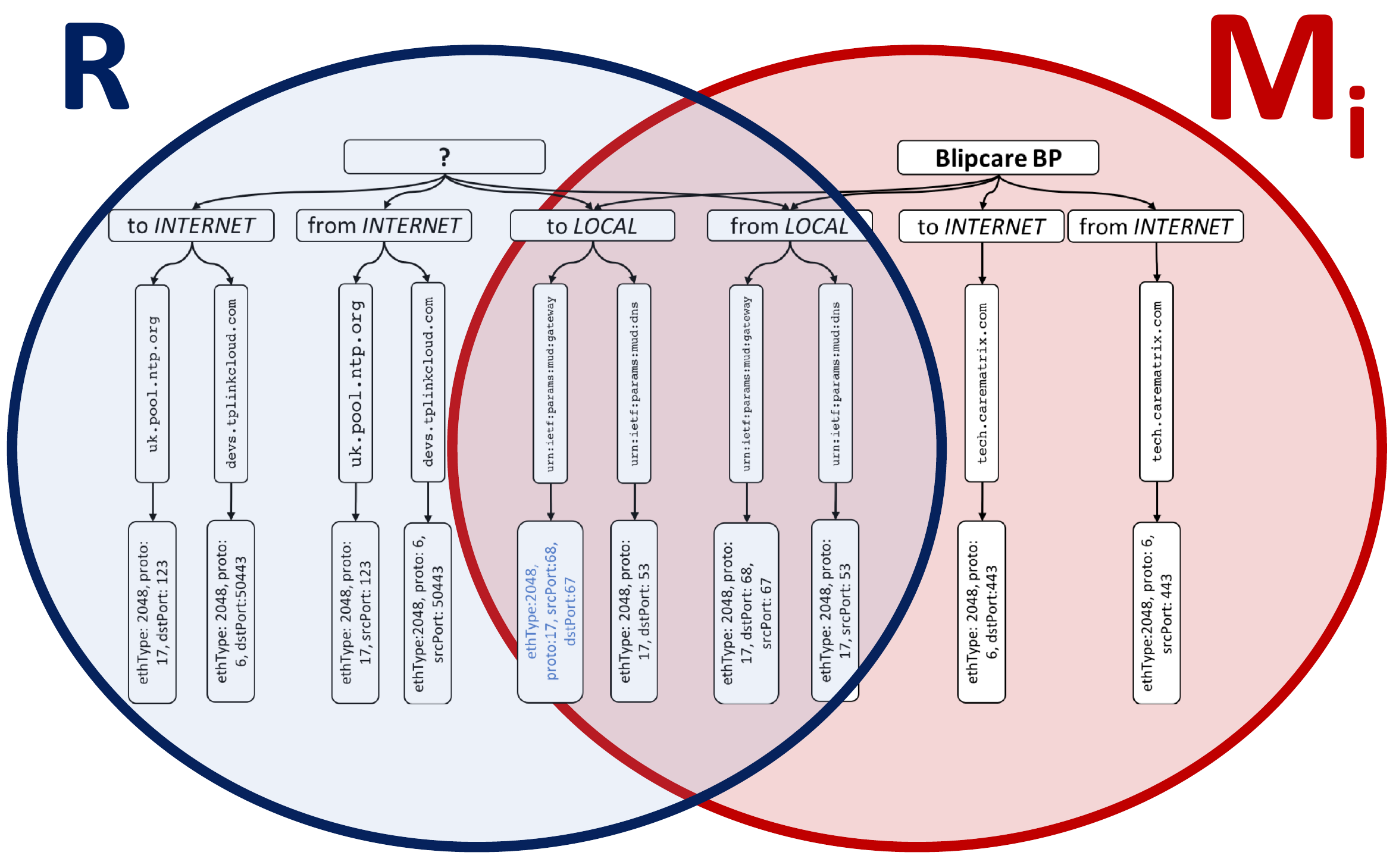}
	\vspace{-3mm}	
	\caption{Comparison of a device's run-time profile $R$ against a known MUD profile $M_i$.}
	\label{fig:check}
	\vspace{-4mm}
\end{figure}

\textbf{Profile structure:}
A device's run-time profile has two key components namely ``Internet'' and ``Local'' communication channels as shown by purple and green regions in Fig.~\ref{fig:tplinkTree}. 
Each profile is organized into a tree-like structure containing a set of nodes with categorical attributes (\ie end-point, protocol, port number over Internet/Local channels) connected through edges. Following the root node in this tree, we have nodes representing the channel/direction of communication, endpoints with which the device communicates, and the flow characteristics (\ie the leaf node). We generate a device's run-time profile as described in \S\ref{sec:mud} with slight variations.

\textit{MUDgee} requires to track the traffic volumes exchanged in each direction for UDP flows to distinguish the UDP server and the client. This can lead to a high consumption of memory when generating run-time profiles. Hence, given a UDP flow, we search all known MUD profiles for an overlapping region. If an overlapping region is found, the tree structure is updated with intersecting port ranges -- this can be seen in Fig.~\ref{fig:tplinkTree} where the leaf node, shown in light-blue text, has been changed according to known MUD profiles. If no overlap is found, we split the UDP flow into two leaf  nodes -- one matches the UDP source port (with a wild-carded destination) and the other matches the UDP destination port (with a wild-carded source). This helps us to identify the server side by subsequent packet matching on either of these flows. 


\textbf{Metrics:}
We denote each run-time profile and MUD profile by the sets $R$ and $M_{i}$ respectively, as shown in Fig.~\ref{fig:check}. 
An element of each set is represented by a branch of the tree structure shown in Fig.~\ref{fig:tplinkTree}. 
For a given IoT device, we need to check the similarity of its $R$ with a number of known $M_{i}$'s. 

There are a number of metrics for measuring the similarity of two sets. \textit{Jaccard index} is widely used for comparing two sets of categorical values, and defined by the ratio of the size of the intersection of two sets to the size of their union, \ie ${|R~\cap~M{i}|}/{|R~\cup~M{i}|}$. Inspired by the Jaccard index, we define the following two metrics:
\begin{itemize}
	\item \textbf{Dynamic similarity score:} $sim_{d}(R,M_{i})=\frac{|R~\cap~M_{i}|} {|R|}$
	\item \textbf{Static similarity score:} $sim_{s}(R,M_{i})=\frac{|R~\cap~M_{i}|} {|M_{i}|}$
\end{itemize}
These two metrics collectively represent the Jaccard index.
Each metric can take a value between $0$ (\ie disjoint) and $1$ (\ie identical). Similarity scores are computed per epoch (\eg 15 minutes). 
When computing $|R~\cap~M{i}|$, we temporarily morph the run-time profile based on each MUD profile it is checked against. This assures that duplicate elements are pruned from $R$ when checking against each $M_{i}$.


\begin{figure*}[!t]
	\begin{center}
		\mbox{
			\subfigure[Static similarity score.]{
				{\includegraphics[width=0.30\textwidth]{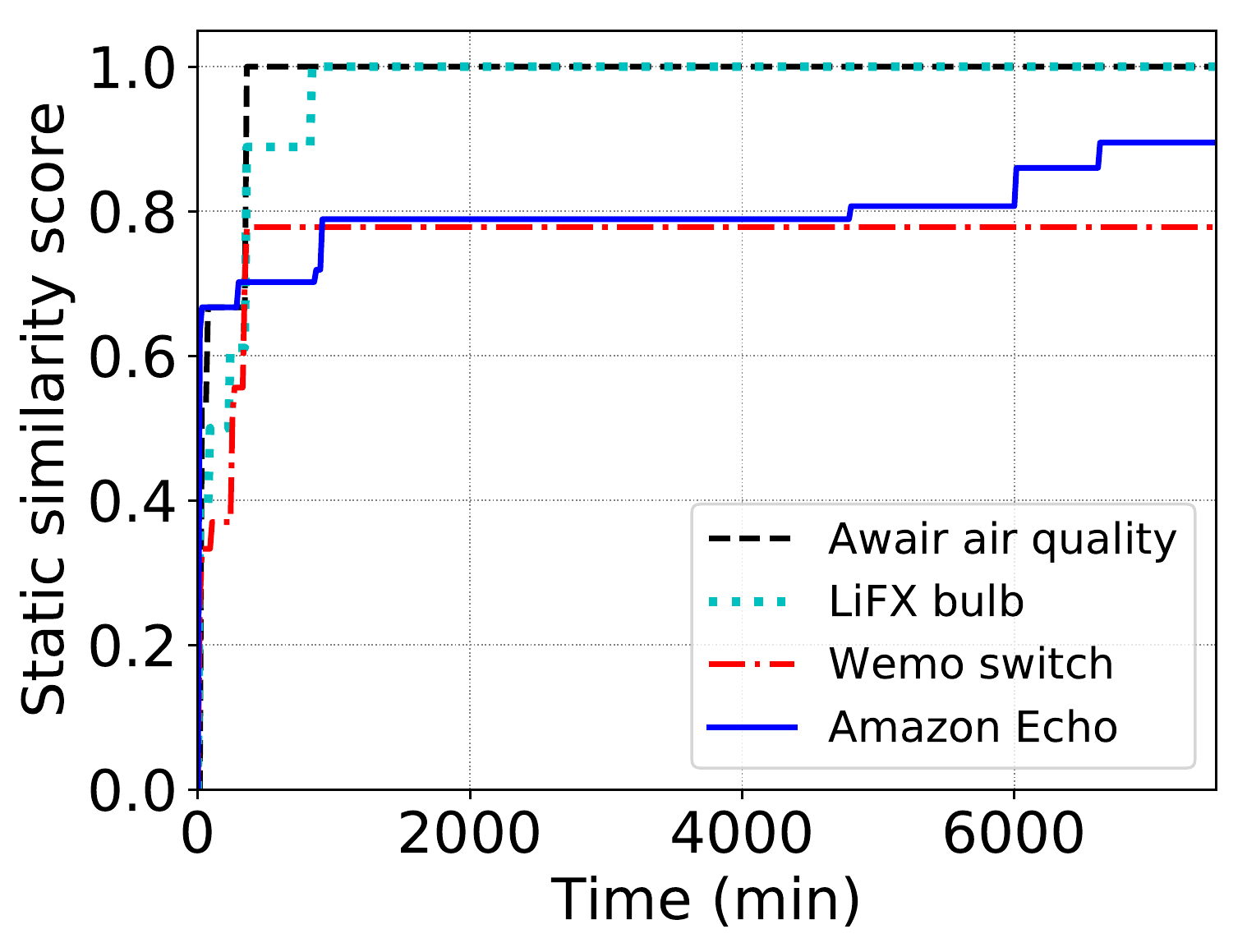}}\quad
				\label{fig:staticsimilarity}
			}
			\hspace{-3mm}
			\subfigure[Dynamic similarity score.]{
				{\includegraphics[width=0.30\textwidth]{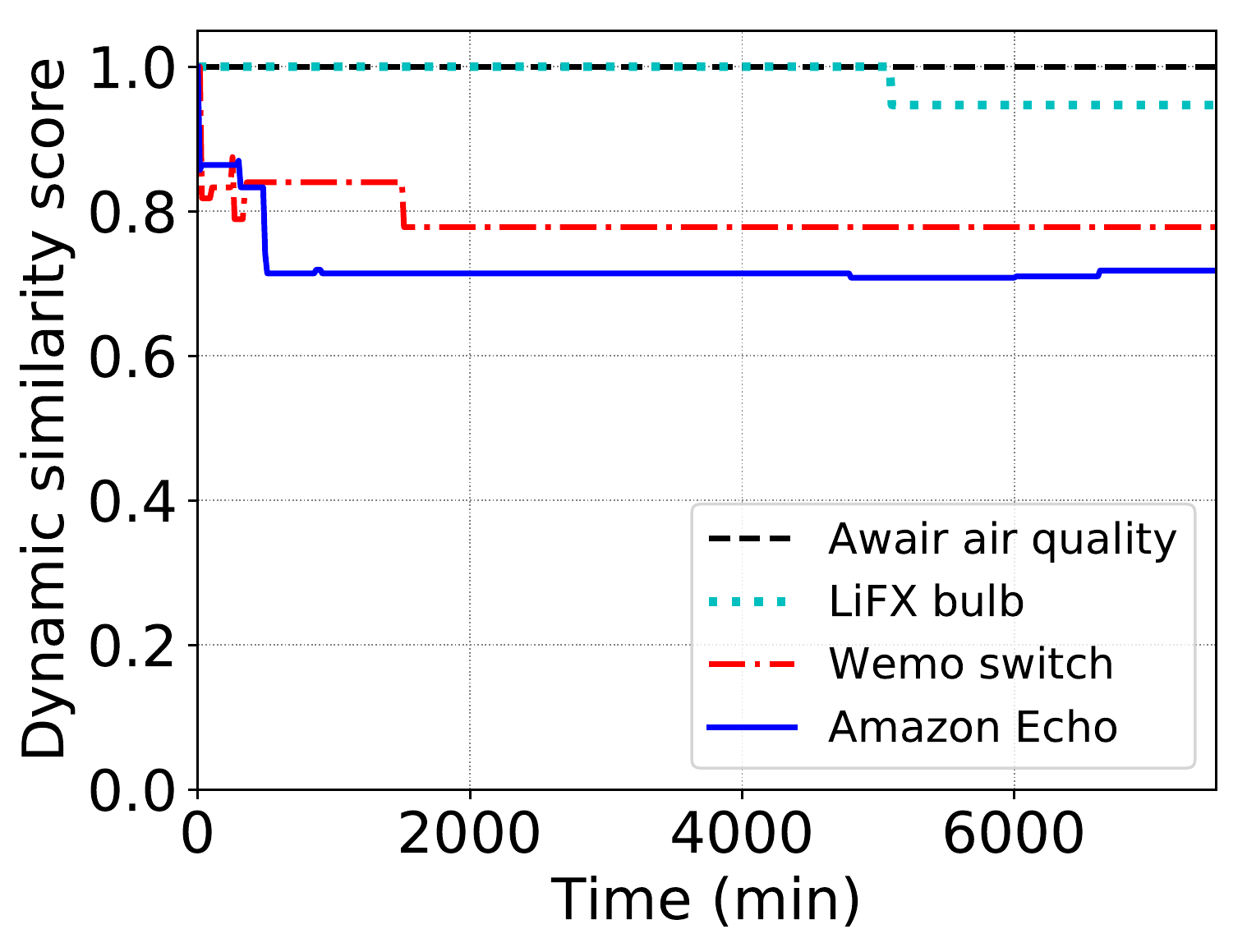}}\quad
				\label{fig:dynamicsimilarity}
			}
			\hspace{-3mm}
			\subfigure[Dynamic similarity score (SSDP excluded).]{
				{\includegraphics[width=0.30\textwidth]{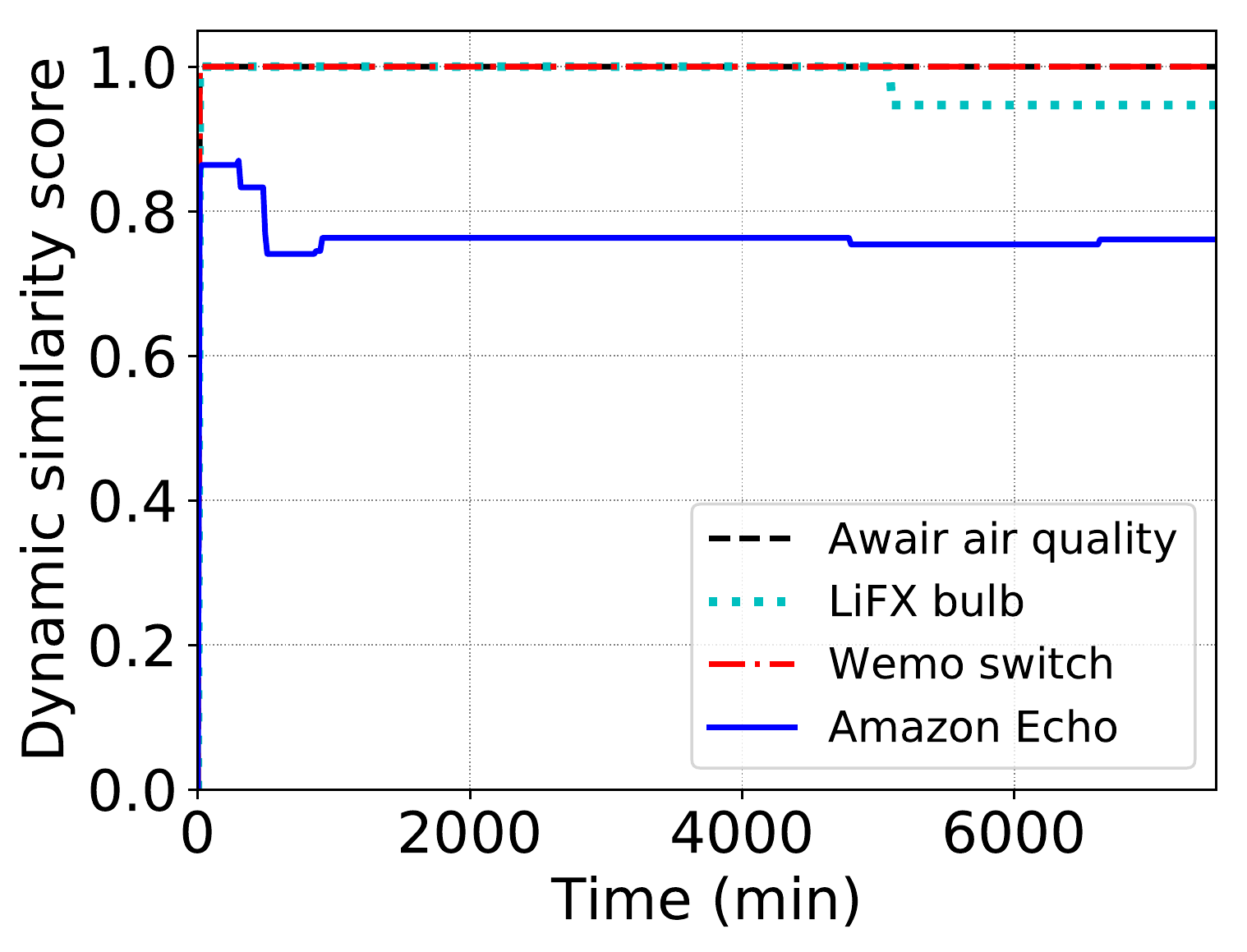}}\quad
				\label{fig:dynamicsimilarityExSSDP}
			}		
		}
		\vspace{-2mm}
		\caption{Time-trace of dynamic and static similarity scores for the winners of four IoT devices. Convergence time depends on the behaviour complexity of the device; for example, the static similarity score of the LiFX bulb converges to $1$ within 1000 minutes whereas it takes about 12 days for the more complex Amazon echo to converge.}
		\vspace{-3mm}
		\label{fig:similarityTimeTrace}
	\end{center}
\end{figure*}

We note that the run-time profile grows over time by accumulating nodes (and edges), as shown by the example in Fig.~\ref{fig:tplinkTree}.
As per the figure, 30 minutes into profile generation, the run-time profile of the TP-Link power plug consists of eight elements (\ie edges).
This element count reaches 15 when additional device traffic is processed (Fig.~\ref{fig:tplinkTree480min}). 

At the end of each epoch, a device (or a group of devices) with the maximum similarity score will be chosen as the ``winner''.
We expect to find a group of devices as the winner when considering dynamic similarity,
because only a small subset of the device's behavioral profile is observed initially.
The number of winners will reduce as the device's run-time profile grows over time.

Fig.~\ref{fig:similarityTimeTrace} shows the time trace of similarity scores for the winners Awair air quality, LiFX bulb, WeMo switch, and Amazon Echo. 
In each plot, a single correct winner is identified per device.
As Fig.~\ref{fig:staticsimilarity} shows, the static similarity score grows slowly over time in a non-decreasing fashion. 
The convergence time depends on the complexity of the device's behavioral profile. 
For example, the static similarity score of Awair air quality and LiFX bulb converges to $1$ within 1000 minutes. 
But for the Amazon Echo, it takes more time to gradually discover all flows -- the convergence time is about 12 days. 

There are also devices for which the static similarity may not converge to $1$.
For example, WeMo switch and WeMo motion use a list of hard-coded IP addresses (instead of domain names) for their NTP communications. 
These IP addresses are now obsolete; no NTP reply flows are captured. 
Likewise, the TPLink plug uses the domain ``{\myverb{s1b.time.edu.cn}}'' for NTP communication
and this domain is also no longer operational. 
Devices such as the August doorbell and Dropcam also contact public DNS resolvers (\eg {\myverb{8.8.4.4}}) 
if the local gateway fails to respond to a DNS query from the IoT device. 
This specific flow can only be captured if there is an Internet outage. 

On the other hand, the dynamic similarity score grows quickly as shown in Fig.~\ref{fig:dynamicsimilarity}.
It may even reach $1$ (\ie $R \subset M_{i}$) and stay at $1$, if no deviation is observed -- deviation is the complement of the dynamic similarity measured in the range of $[0,1]$ and computed as $1 - sim_{d}$ . 
Awair air quality monitor exhibits such behavior as shown by the dashed black lines in Fig.~\ref{fig:dynamicsimilarity}
-- 19 out of 28 IoT devices in our testbed exhibit similar behavior in their dynamic similarity scores. 
In other cases, these scores may fluctuate;
a fluctuating dynamic similarity never meets $1$ due to missing elements (\ie deviation). 
Missing elements can arise due to 
(a) a MUD profile being unknown or not well-defined by the manufacturer, 
(b) a device firmware being outdated, 
and (c) an IoT device being compromised or under cyber attack.

\begin{figure}[t!]
	\centering
	\vspace{-5mm}		
	\includegraphics[scale=0.395]{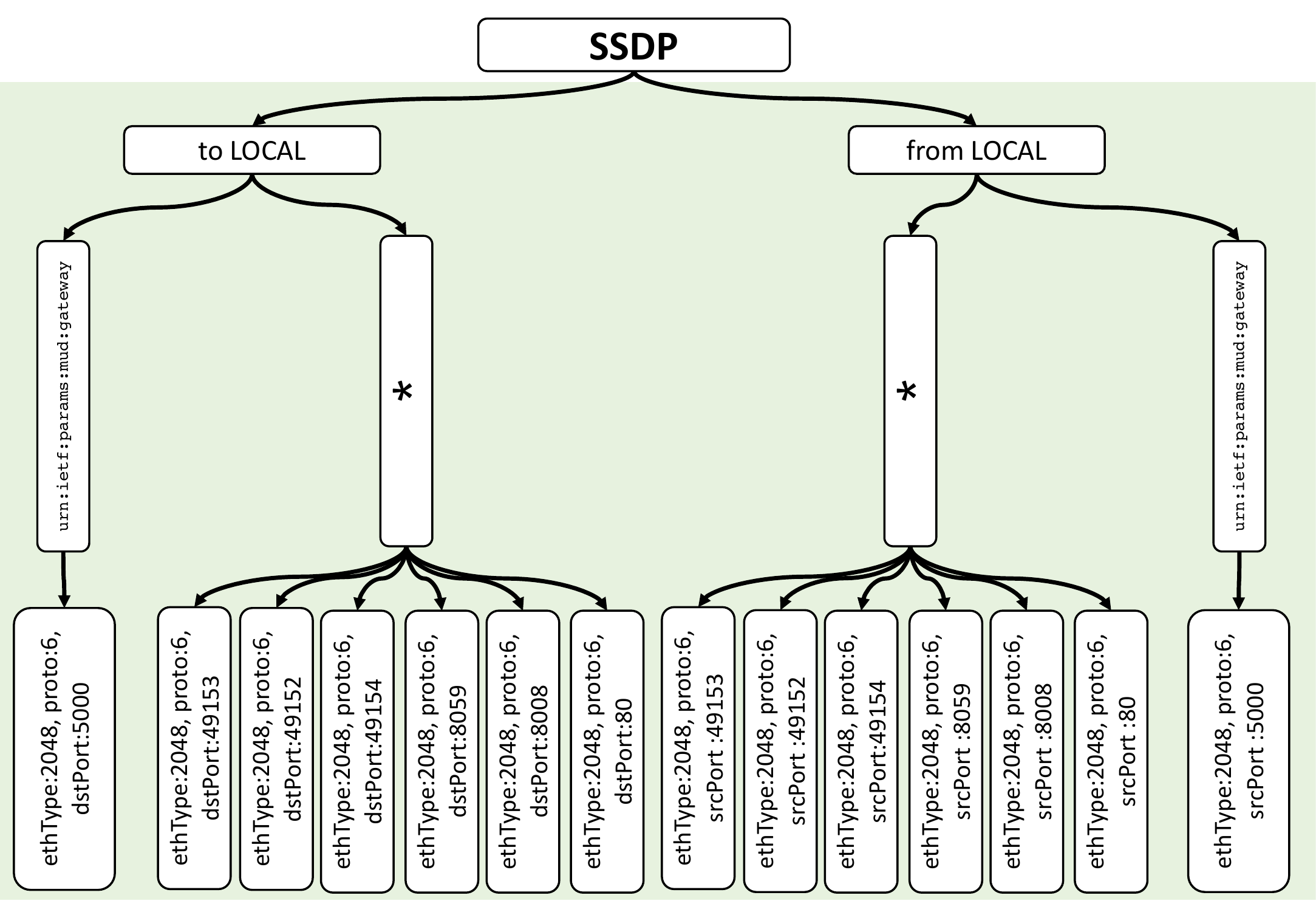}
	\vspace{-3mm}	
	\caption{SSDP runtime profile describing all discovery communications across all devices in the network.}
	\label{fig:ssdpProfile}
	\vspace{-4mm}
\end{figure}

We found that nine of our testbed IoTs had slight deviation. These were due to two reasons,
Firstly, when responding to discovery requests in Local communications; if the devices support the SSDP protocol \footnote{Devices that supports Simple Service discovery protocol advertises or notify device capabilities to Multicast UDP port 1900. Typically the payload contains device information including IP address, name, UUID, management URL, functionalities.}, these responses cannot be tightly specified by the manufacturer in the MUD profile as such flows depend on the environment in which the IoT device is deployed. An example is the WeMo switch, shown by dashed-dotted red lines  in Fig.~\ref{fig:dynamicsimilarity}. 
We populate all discovery communications in a separate profile (shown in Fig.~\ref{fig:ssdpProfile}) by inspecting SSDP packets exchanged over the local network to address this issue. 
We note that the SSDP server port number on the device can change dynamically, thus the inspection of the first packet in a new SSDP flow is required.
The second reason for deviation is missing DNS packets which can lead to emergence of a branch in the profile with an IP address as the end-point instead of a domain name. 
This can occur in our testbed because each midnight we start storing traffic traces into a new PCAP file, thus few packets may get lost during this transition. 
Missing DNS packets were observed for the LiFX bulb, as shown by dotted cyan lines in Fig.~\ref{fig:dynamicsimilarity}.   

Thus, we exclude SSDP activity from local communications of IoT devices to obtain a clear run-time profile.
As Fig.~\ref{fig:dynamicsimilarityExSSDP} shows, the filtering allows 
us to correctly identify the winner for the WeMo switch within a very short time using 
the dynamic similarity score.

\begin{figure}[t!]
	\centering
	\vspace{-3mm}		
	\includegraphics[scale=0.45]{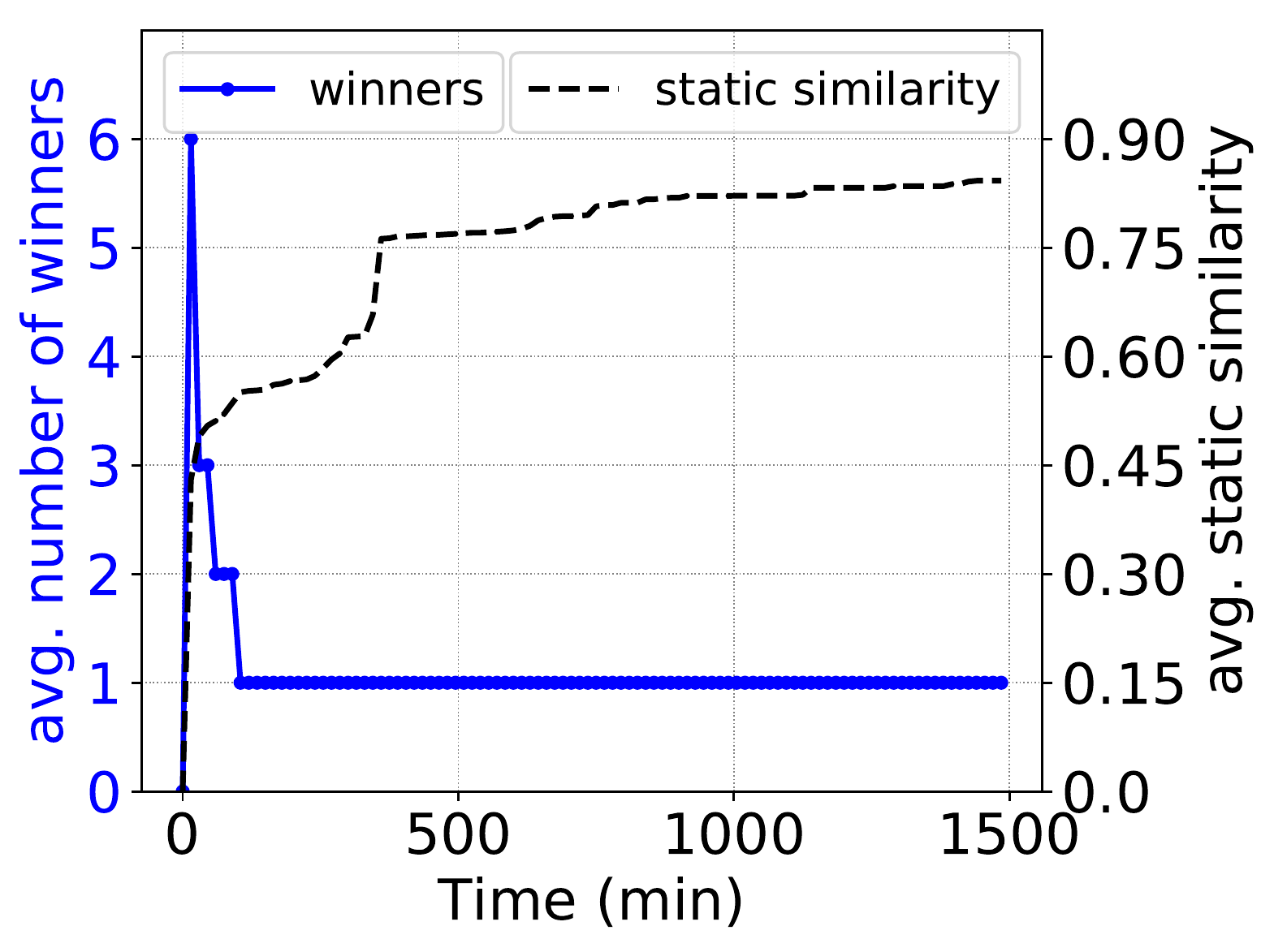}
	\vspace{-4mm}	
	\caption{Time trace of winners count and static similarity score averaged across 27 testbed IoT devices. The former shows six winners on average at the beginning of the identification process. This count drops to a single winner in less than three hours. Even with a single winner, the static similarity needs about ten hours on average to exceed a threshold of $0.8$.}
	\label{fig:numwinnersAndSimScore}
	\vspace{-4mm}
\end{figure}

\begin{figure*}[t!]
	\centering
	\includegraphics[width=0.90\textwidth]{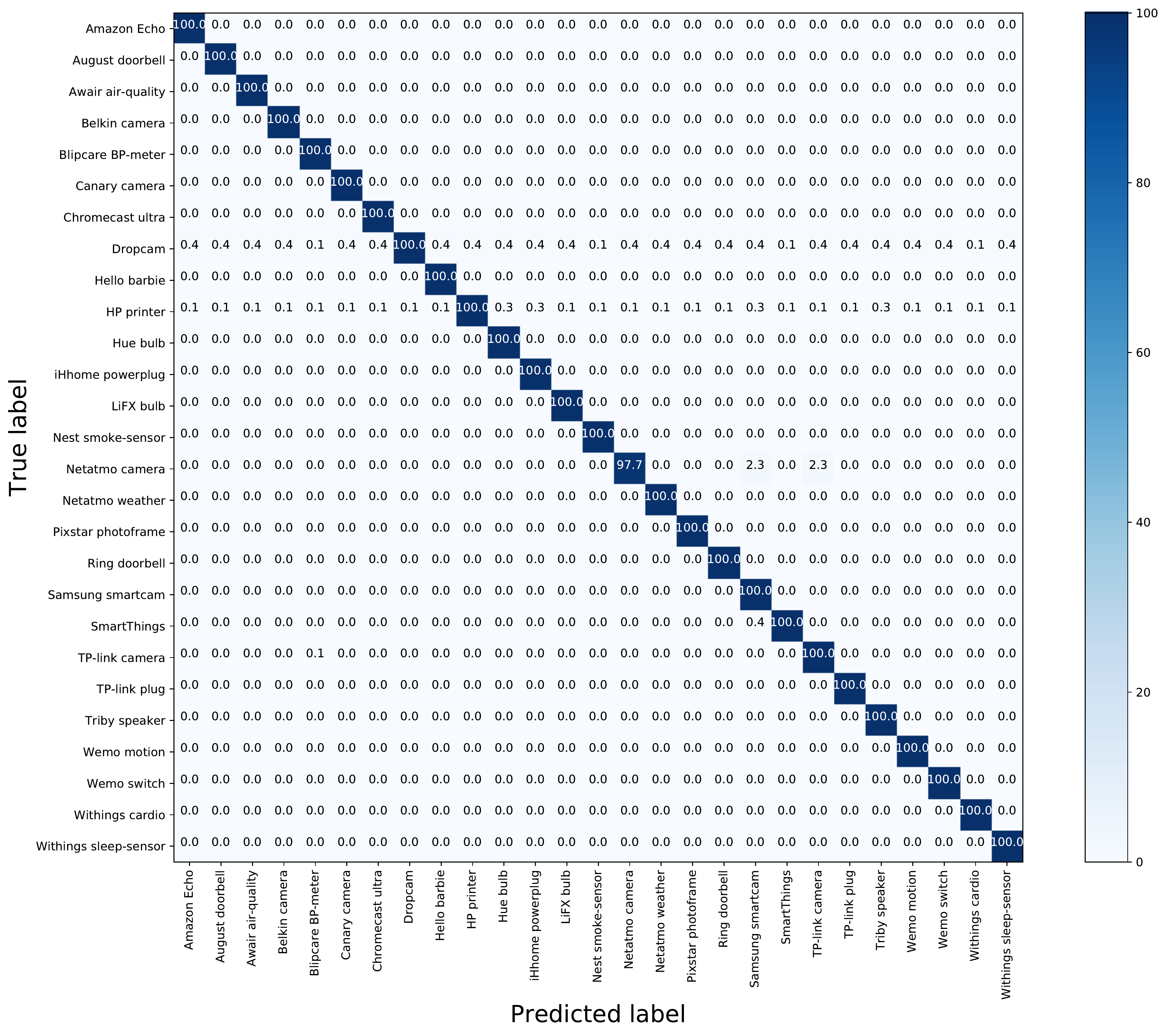}
	\vspace{-4mm}	
	\caption{Confusion matrix of true vs predicted device labels. The cell values are in percentages. As the table shows, for instance, the Amazon Echo (first row) is always predicted as the sole winner in all epochs. Hence, a value of $100$\% is recorded in the first column and $0$\% in the remaining columns.}
	\label{fig:confMapValid}
	\vspace{-4mm}
\end{figure*}

Lastly, it is important to note that similarity scores (both static and dynamic) can be computed at an aggregate level (\ie Local and Internet combined), or per individual channel.
The latter may not converge in some cases where the Local channel similarity finds one winner
while the Internet channel similarity finds a different winner. We note that per-channel similarity never results in a wrong winner, but may result in finding no winners. 
In contrast, aggregate similarity can lead to the wrong winner, especially when Local activity becomes dominant in the behavioral profile. This is because many IoTs have a significant profile overlap in their Local communications (\eg DHCP, ARP, or SSDP). 
Hence, we begin by checking per-channel similarity, if the two channels disagree, we switch to aggregate similarity to identify the winner. 
We discuss this scenario in detail in \S\ref{sec:changeMUD}.

%

\vspace{-3mm}
\subsection{Identifying IoT Devices at Run-Time} 
\textbf{Dataset:} We use packet traces (\ie {\myverb{PCAP}} files) collected from our testbed comprising a gateway (\ie a TP-Link Archer C7 flashed with the OpenWrt firmware) that serves a number of IoT devices. We store all network traffic (Local and Internet) onto a 1TB USB storage connected to this gateway using {\myverb{tcpdump}}. Our traffic traces span three months, starting from May 2018, containing traffic corresponding to devices listed in Table~\ref{table:IoTdevices} (excluding Withings baby monitor). 
We used \textit{MUDgee} to generate the MUD profiles for these devices. 
We also developed an application over our native SDN simulator \cite{sdnsimulator} to implement our identification process.

\begin{figure*}[t]
	\centering
	\includegraphics[width=0.65\textwidth]{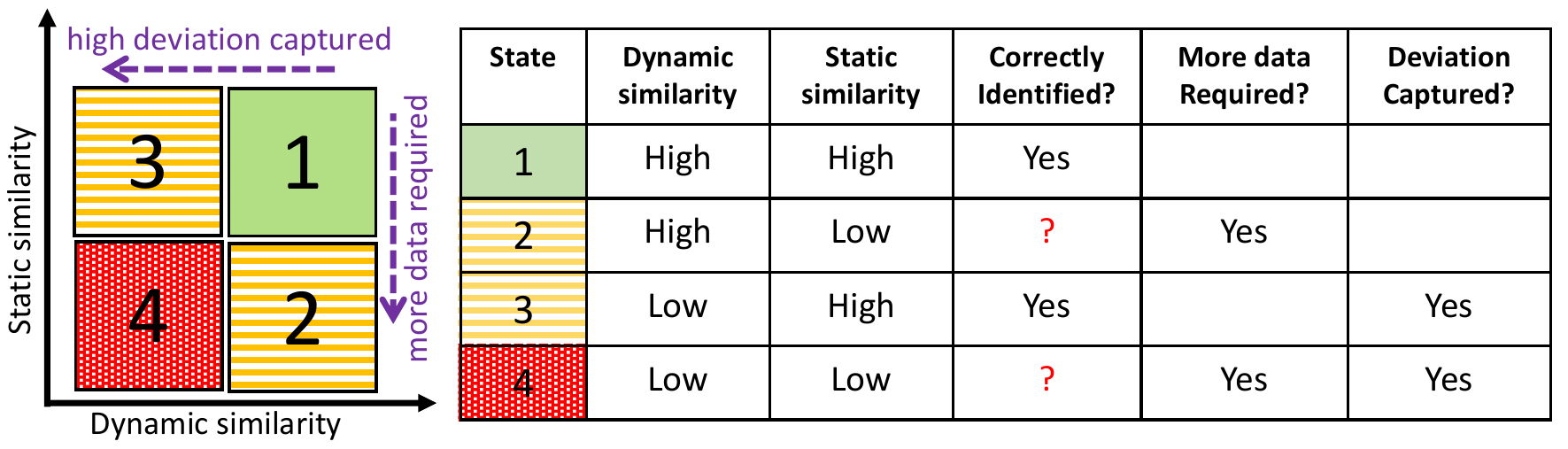}
	\vspace{-3mm}	
	\caption{Plot of dynamic similarity vs static similarity depicting 4 distinct states. In state-1, both dynamic and static similarity scores are high and  we obtain a single correct winner.
		In state-2, dynamic similarity is high but static similarity is low (usually occurs when only a small amount of traffic is observed). State-3 describes a region with high static similarity yet
		low dynamic similarity, indicating high-deviation at run time (\eg due to old firmware or device being compromised). In state-4 both dynamic and static similarity scores are low indicating a significant difference between the run-time and MUD profiles.}
	\label{fig:staticVsDynamic}
\end{figure*}

\begin{figure*}[t!]
	\centering
	\includegraphics[width=0.98\textwidth]{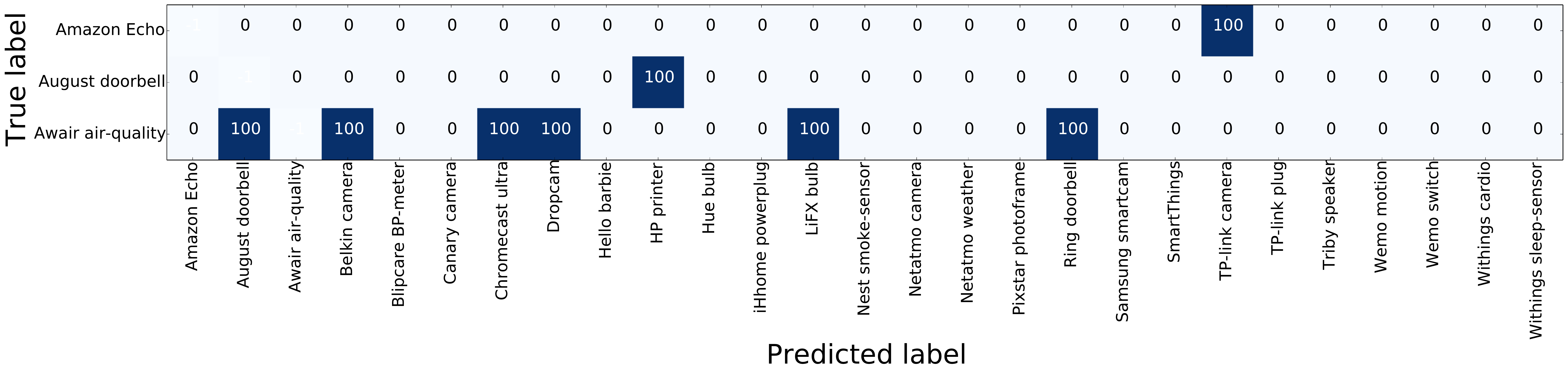}
	\vspace{-4mm}	
	\caption{Partial confusion matrix for when the intended MUD profile is absent for each device being checked.}
	\label{fig:inverseconfMapValid}
	\vspace{-2mm}
\end{figure*}

\begin{figure*}[t!]
	\begin{center}
		\mbox{
			\subfigure[Dynamic similarity score.]{
				{\includegraphics[width=0.35\textwidth]{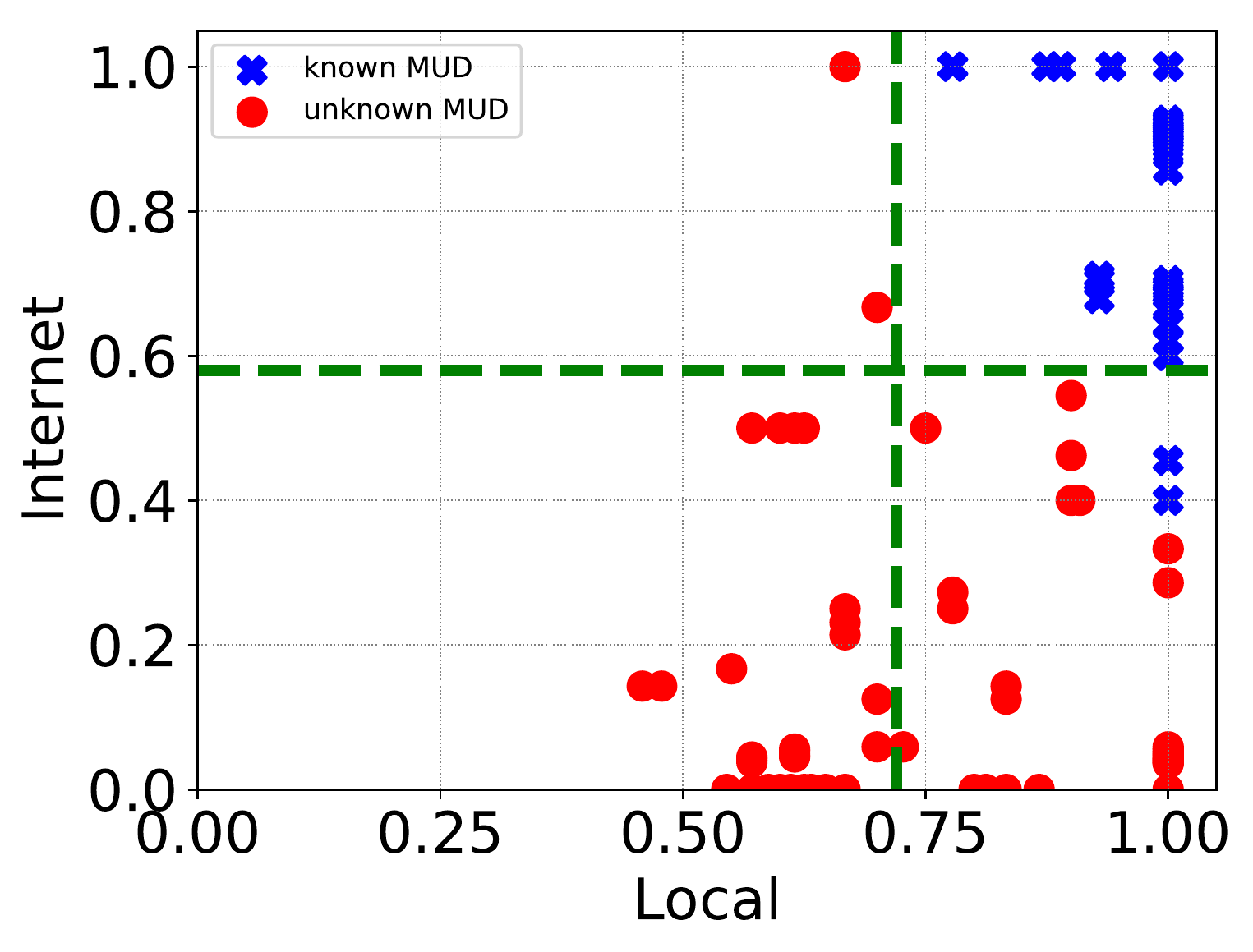}}\quad
				\label{fig:dynamicsimilarityThr}
			}	
			\hspace{-3mm}
			\subfigure[Static similarity score.]{
				{\includegraphics[width=0.35\textwidth]{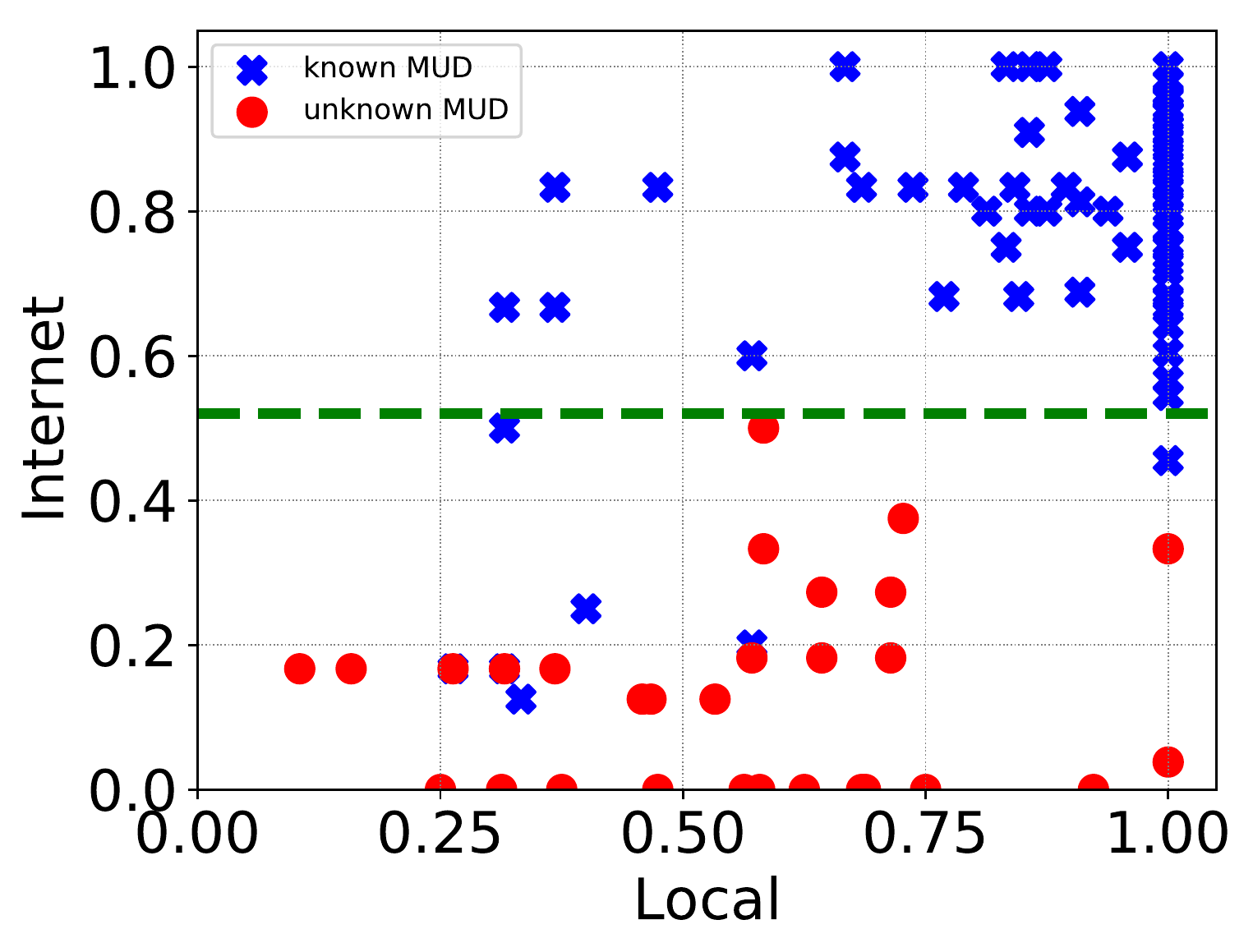}}\quad
				\label{fig:staticsimilarityThr}
			}			
		}
		\vspace{-2mm}
		\caption{Scatter plots of channel-level scores for dynamic and static similarity metrics across 27 testbed IoT devices.
			Each plot depicts two sets of results: one for known MUD (blue markers) and the other for unknown MUD (red markers).
			Enforcing two thresholds (\ie about 0.60 on the Internet channel and 0.75 on the Local channel) would filter incorrect matches found using dynamic similarity.
			A threshold of 0.50 on the Internet channel is sufficient to avoid false identification when using static similarity.}
		\vspace{-6mm}
		\label{fig:similairityScoreComparison}
	\end{center}
\end{figure*}

\textbf{Identification Process:} As described earlier, a dynamic similarity score converges faster than a static similarity score. So, our device identification process begins by tracking dynamic similarity at a channel level and continues as long as channel agreement persists. Depending on the diversity of observed traffic to/from the IoT device (Local vs Internet), there may be multiple winners at the beginning of this process. At this point, static similarity is fairly low, since only a small fraction of the expected profile is likely captured in the short period. Hence, our process needs additional traffic as input for the device before it can conclude winners.
Fig.~\ref{fig:numwinnersAndSimScore} shows the time-trace evolution of winners count with static similarity, averaged across our 27 testbed IoT devices. 
The solid blue line (left y-axis), shows up to six winners on average at the beginning of the identification process. 
This count gradually drops (in less than three hours) to a single winner and
stabilizes.
Even with a single winner, the static similarity, shown by dashed black lines (right y-axis), needs about ten hours on average to pass a threshold score of $0.8$. 
Reaching a static similarity score of $1$ can take long (a full score may also not be reached). So, the network operator must choose
an appropriate threshold to conclude traffic processing -- a higher threshold increases the device identification confidence level, but comes at a cost of longer convergence time.

We replayed our packet traces collected in 2018 (\ie Data-2018) into our packet simulator tool.
Fig.~\ref{fig:confMapValid} shows a confusion matrix of the results -- rows are actual device labels, columns are predicted device labels, and cell values are in percentage. 

The table depicts the efficacy of our approach; for example, the first row in the table shows that the Amazon Echo is always predicted as the sole winner in each epoch.
Hence, a value $100$\% is recorded in the first column and $0$\% in the remaining columns.
No other device is identified as the winner in any epoch.
Considering the row containing Dropcam, the device is identified as another in some epochs.
Hence, non-zero values are recorded against all columns. 
But, Dropcam is always one of the winners, \ie its column records a value of $100$\%.

We observe correct convergence for all devices except for the Netatmo camera where it is not correctly identified in $2.3$\% of epochs. 
This mis-identification occurs due to missing DNS packets where some flows are incorrectly matched on STUN related flows (with wild-carded endpoints) of Samsung camera and TP-Link camera. This mis-identification occurs only during the first few epochs, the process subsequently converges to the correct winner.  
In what follows, we discuss changes in IoT traffic behaviour in the network.

\vspace{-3mm}
\subsection{Monitoring Behavioral Change of IoTs}\label{sec:changeMUD}

In practice, identifying an IoT device at runtime gives rise to several challenges: 
(a) the network device may not have a known MUD profile, 
(b) the device firmware may be outdated (thus, the run-time profile can deviate from its current MUD profile), and 
(c) the device may be under attack or compromised. We focus on these issues here and discuss our methodology to addressing these challenges.


Fig.~\ref{fig:staticVsDynamic} depicts a simplified scatter plot of dynamic similarity versus static similarity, 
In this plot, there are color-coded states labeled 1, 2, 3, and 4.
Our ideal region is the green quadrant (\ie state-1) where both dynamic and static scores are high, and we have a single correctly identified winner.
State-2 describes a region with a high dynamic similarity score and a fairly low static similarity score. 
We expect this state when only a small amount of traffic from the device is observed and 
additional traffic is needed to evaluate if dynamic similarity will continue to remain high and
static similarity starts rising.
State-3 describes a region with high static similarity yet low dynamic similarity -- this is indicative of high deviation at run-time.
We observe this state when many flows identified in actual device traffic are not listed in the intended MUD profile.
This can be due to two reasons:   
(a) the device firmware not being current, or 
(b) the device being under attack or compromised.
Finally, having low dynamic and static similarity scores highlight a significant difference
between the run-time and MUD profiles. This scenario likely results in an incorrectly identified winner. 

In summary, IoT network operators may need to set threshold values for both dynamic and static similarity scores to select a winner device. 
The identification process must also
begin with channel-level similarity (for both dynamic and static scores) 
and switch to aggregate-level in case of non-convergence.
In what follows, we quantity the impact of three scenarios enabling IoT behavioral changes:

\noindent \textit{\textbf{MUD profile unknown:}} We begin by removing a single MUD profile at a time from a list of known MUD signatures.
Fig.~\ref{fig:inverseconfMapValid} shows the partial results for each selected device. 
Unsurprisingly, each row device is identified as another (\ie wrong winner selected) 
since its intended MUD profile is absent.
For example, Amazon Echo converges to TP-Link camera and Awair air quality monitor is consistently identified as six other IoTs. 
Ideally, we should have no device identified as a winner.
It is important to note here, that
these results were derived without applying thresholds to the similarity scores - \ie
only the maximum score was used to pick winners.


Fig.~\ref{fig:similairityScoreComparison} shows scatter plots of channel-level scores for both dynamic and static similarity metrics across our testbed IoT devices.
In each plot we depict two sets of results generated using our Dataset-2018:
one for known MUD (shown by blue cross markers) and the other for unknown MUD (shown by red circle markers). 
Enforcing two thresholds (\ie about $0.60$ on the Internet channel and $0.75$ on the Local channel) would filter incorrect matches
found using dynamic similarity (\ie Fig.~\ref{fig:dynamicsimilarityThr}).
A threshold of $0.50$ on the Internet channel is sufficient to avoid incorrect identification when using static similarity (Fig.~\ref{fig:staticsimilarityThr}).
A single threshold is sufficient for the latter because device behaviour on the Internet channel varies significantly 
for the consumer devices we have running in our testbed,
but enterprise IoTs may tend to be more active on the Local network, requiring a different thresholding mechanism.

We note here that a high threshold value increases the time to identification and a low threshold value reduces it, but can also lead to an incorrect winner. 
Hence, it is up to the network operator to set threshold values. 
A conservative approach may accept no deviation in dynamic similarity with a static similarity score over $0.50$ 
per Local and Internet channel.

We regenerated the results using these conservative thresholds and found there were no winners due to low scores in both dynamic and static-similarity metrics.
This indicates that devices, in the absence of their MUD profiles, are consistently found in state-4 in Fig.~\ref{fig:staticVsDynamic}, flagging possible issues.

\noindent \textit{\textbf{Old firmware:}} IoT devices usually upgrade their firmware automatically by directly communicating with a cloud server, or require the user to confirm the upgrade (\eg WeMo switch) via an App. 
In the latter case, devices can stay behind the latest firmware until the user manually updates them.
To illustrate the impact of old firmware, we used packet traces collected from our testbed for a duration of six months starting in October 2016. 
We replayed Data-2016 to check run-time profiles against the MUD profiles generated from Data-2018.
Table~\ref{table:firmwareResult} shows the results. The column labeled ``Profile changed'' indicates whether any changes on device's behavior is observed from Data-2016 compared with Data-2018. These behavioral changes include endpoints and/or port numbers. For example, TP-Link camera  communicates with a server endpoint ``{\myverb{devs.tplinkcloud.com}}'' on TCP 50443 as per Data-2016. However, this camera communicates with the same endpoint on TCP 443 as per Data-2018. 
Additionally, as per this dataset, an endpoint ``{\myverb{ipcserv.tplinkcloud.com}}'' is observed which did not exist in Data-2016.  




\begin{figure}[t]
	\centering
	\vspace{-6mm}	
	\includegraphics[width=0.48\textwidth]{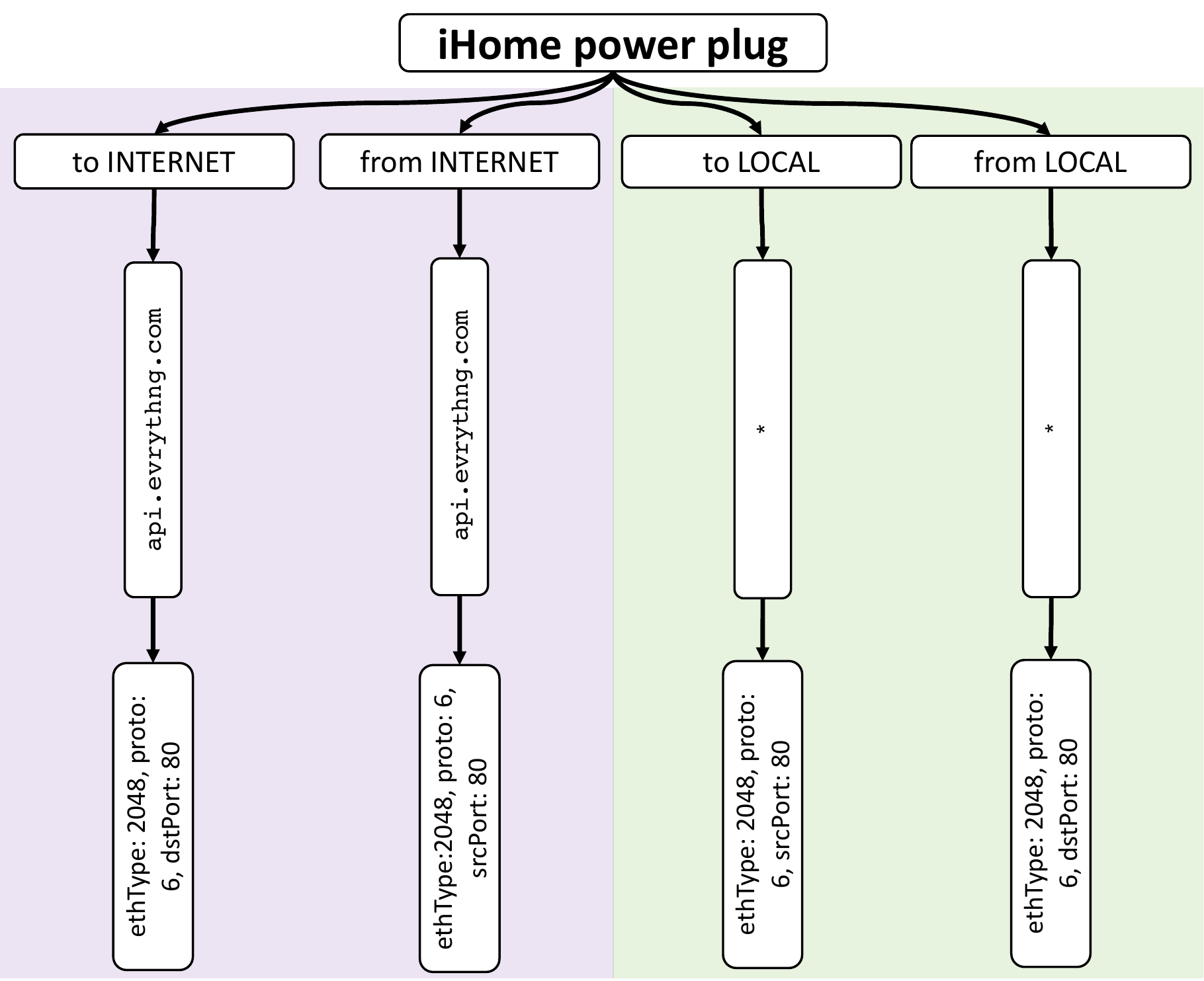}
	\vspace{-2mm}	
	\caption{Tree structure depicting profile difference (\ie $R$ - $M$) for the iHome power plug.}
	\label{fig:firmwareprofile}
	\vspace{-4mm}
\end{figure}

\begin{table*}
	\centering
	\caption{Identification results for data 2016.}\label{table:firmwareResult}
	\vspace{-2mm}
	\begin{adjustbox}{max width=0.88\textwidth}	
		\begin{tabular}{|l|c|p{1.2cm}|p{1.2cm}|p{1.2cm}|p{1.2cm}|p{1.2cm}|p{1.2cm}|p{1.2cm}|p{1.2cm}|p{1.2cm}|p{1.2cm}|}
			\hline 
			\multirow{3}{*}{\textbf{IoT device}} & \multirow{3}{*}{\rotatebox{90}{Profile change}} & \multicolumn{3}{c|}{\textbf{Convergence}} & \multicolumn{4}{c|}{\textbf{Convergence with threshold}} & \multicolumn{3}{c|}{\textbf{Endpoint compacted}}\tabularnewline
			\cline{3-12} 
			&& \multicolumn{2}{c|}{Known MUD} & Unknown MUD & \multicolumn{3}{c|}{Known MUD} & Unknown MUD  & \multicolumn{2}{c|}{Known MUD} & Unknown MUD\tabularnewline
			\cline{3-12}
			&& Correctly  identified (\%) & Incorrectly identified (\%) & Incorrectly identified (\%) & Correctly identified (\%) & Incorrectly identified (\%) & State & Incorrectly identified (\%) & Correctly identified (\%) & Incorrectly identified (\%) & Incorrectly identified (\%)\tabularnewline
			\hline 
			Amazon Echo& Yes & 100 & 0 & 100 & 65.7 & 0 & 3 & 0& 65.7 & 0 & 0\tabularnewline
			\hline 
			August doorbell & Yes & 100 & 0 & 100 & 0 & 0 & 4 & 0 & 100 & 0 & 0\tabularnewline
			\hline 
			Awair air quality& Yes & 100 & 0 & 100 & 100 & 0  & 1 & 0& 100 & 0 & 0\tabularnewline
			\hline 
			Belkin camera& Yes & 100 & 0 & 100 & 100 & 0 &  1 &0 & 100 & 0 & 0\tabularnewline
			\hline 
			Blipcare BP meter& No & 100 & 0 & 100 & 100 & 0 &  1 &0 & 100 & 0 & 0\tabularnewline
			\hline 
			Canary camera& No & 100 & 0 & 100 & 100 & 0 &  1 & 0 & 100 & 0 & 0\tabularnewline
			\hline 
			Dropcam& Yes & 100 & 0 & 100 & 95.9 & 0 &  3 &0 & 100 & 0 & 0\tabularnewline
			\hline 
			Hello barbie& No & 100 & 0 & 100 & 100 & 0 &  1 & 0 & 100 & 0 & 0\tabularnewline
			\hline 
			HP printer& Yes & 100 & 0 & 100 & 3.6 & 0 &  4 &0 & 99.8 & 0 & 0\tabularnewline
			\hline 
			Hue bulb& Yes & 100 & 0 & 100 & 0 & 0 & 4 & 0  & 90.6 & 0 & 0\tabularnewline
			\hline 
			iHome power plug& Yes & 100 & 0 & 100 & 0.5 & 0  & 4& 0 & 100 & 0 & 0\tabularnewline
			\hline 
			LiFX bulb& No & 100 & 0 & 100 & 100 & 0 &  1 &5.3 & 100 & 0 & 5.3\tabularnewline
			\hline 
			Nest smoke sensor& Yes & 100 & 0 & 100 & 0 & 0 &  4 & 0 &100 & 0 & 0\tabularnewline
			\hline 
			Netatmo camera& Yes & 99.4 & 0.6 & 100 & 97.3 & 0 & 3 & 0  & 99 & 0 & 0\tabularnewline
			\hline 
			Netatmo weather& No & 100 & 0 & 100 & 100 & 0  & 1& 0& 100 & 0 & 0\tabularnewline
			\hline 
			Pixstar photoframe& No & 100 & 0 & 100 & 100 & 0  & 1 & 0 & 100 & 0 & 0\tabularnewline
			\hline 
			Ring doorbell& Yes& 100 & 0 & 100 & 99.6 & 0 & 3 & 0 & 97.9 & 0 & 0\tabularnewline
			\hline 
			Samsung smartcam& Yes& 100 & 0 & 100 & 97.6 & 0 & 1 & 0 & 97.6 & 0 & 0\tabularnewline
			\hline 
			Smart Things& No & 100 & 0 & 100 & 100 & 0 & 1 & 0 & 100 & 0 & 0\tabularnewline
			\hline 
			TPlink camera& Yes& 100 & 0 & 100 & 100 & 0 & 3 & 0 & 100 & 0 & 0.9\tabularnewline
			\hline 
			TPlink plug& Yes& 100 & 0 & 100 & 100 & 0 &  1 & 0 & 100 & 0 & 0\tabularnewline
			\hline 
			Triby speaker& Yes& 100 & 0 & 100 & 39.9 & 0  & 3 & 0& 99.8 & 0 & 0\tabularnewline
			\hline 
			WeMo motion& No& 100 & 0 & 100 & 100 & 0  & 1 & 0.7& 100 & 0 & 27.3\tabularnewline
			\hline 
			WeMo switch& Yes& 0 & 100 & 100 & 0 & 100  & 1 & 100& 0 & 100 & 100\tabularnewline
			\hline 
		\end{tabular}
	\end{adjustbox}
\end{table*}

The column ``Convergence" in Table~\ref{table:firmwareResult} describes the performance of our device identification method for two scenarios -- known MUD and unknown MUD. 
When the MUD profile of a device is known, we see that all devices except  theWeMo switch converge to the correct winner. 
Surprisingly, WeMo switch is consistently identified as WeMo motion -- even when its static similarity reaches $0.96$! 
This is because both WeMo motion and WeMo switch share cloud-based endpoints for their Internet communications in Data-2016, 
but these endpoints have changed for the WeMo switch (but not for WeMo motion) in Data-2018.
It is important to note here that our primary objective is to secure IoT devices by enforcing tight access-control rules in policy arbiters.
Therefore, the WeMo switch can still be protected using WeMo motion MUD rules until it gets the latest firmware update. 
Once updated, an intrusion detection system \cite{MUDids2018} may generate false alarms for the WeMo switch,
indicating the need for a re-identification.



As described earlier, we need to enforce thresholds in the identification process to discover unknown devices and resolve problematic states. 
We applied the thresholds determined using Data-2018 and the results are shown in Table~\ref{table:firmwareResult} under ``Convergence with threshold".
Devices without any behavioral changes (from 2016 to 2018), converge correctly and are in state-1. 
In other devices such as the Amazon Echo, only $65.7$\% of instances are correctly identified 
-- the identification process takes considerable time to reach the threshold values.


We observe that devices with profile changes are found in state-3 or state-4. 
These profile differences can be visualised using a tree structure to better understand the causes of a low dynamic similarity score.
Fig.~\ref{fig:firmwareprofile} for instance, shows this difference (\ie $R~-~M$) for the iHome power plug.
As per Data-2016, this device communicates over HTTP with ``{\myverb{api.evrything.com}}" and serves HTTP to the Local network. 
But, these communications do not exist in the MUD profile generated from Data-2018. 
Thus, a firmware upgrade is needed for the device or its current MUD profile is incomplete.


\begin{figure*}[t]
	\centering
	\includegraphics[width=0.80\textwidth]{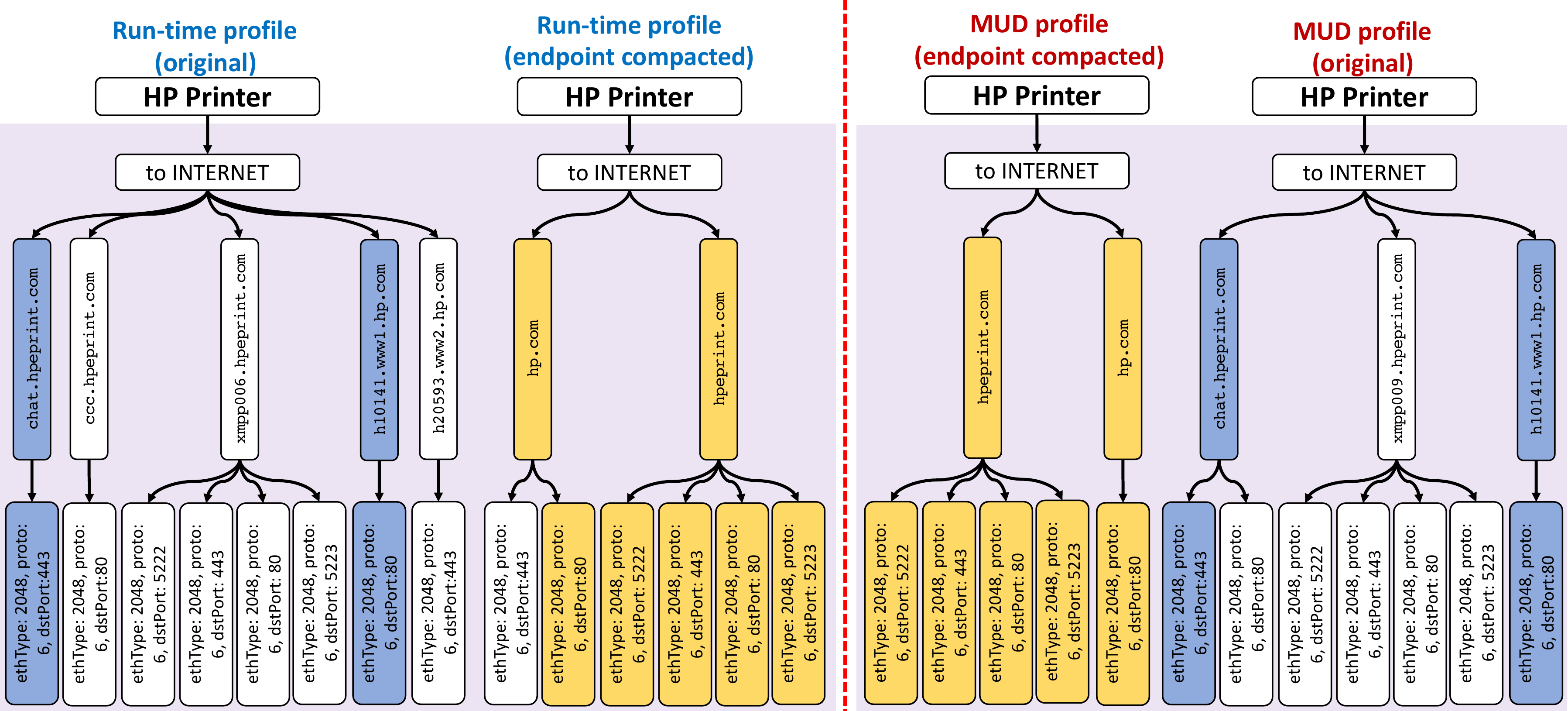}
	\caption{Endpoint compaction of the HP printer run-time and MUD profiles in the ``to Internet" channel direction yields high static and dynamic similarity (shown by the overlapping region in brown). Without compaction these similarities are significantly low (shown by the overlapping region in blue).}
	\label{fig:endpointCompaction}
	\vspace{-4mm}
\end{figure*}

We may find a device (\eg HP printer or Hue bulb) consistently in state-4 throughout the identification process.
Structural deviation in the profile largely arise due to changes in the endpoints or port numbers. 
Tracking port number changes is non-trivial. 
However, for endpoints we can compact fully-qualified domain names to primary domain names (\ie by removing sub-domain names) -- we call this technique as endpoint compaction. 
Note that if the device is under attack or compromised it is likely to communicate with a completely new primary domain. 
Fig.~\ref{fig:endpointCompaction} illustrates endpoint compaction for the HP printer profile in the ``to Internet" channel direction. {\color{blue}}
For this channel, without endpoint compaction, the static and dynamic similarity scores are $0.28$ and $0.25$ respectively. 
Applying endpoint compaction yields much higher similarity scores of $1$ and $0.83$, respectively.

We applied endpoint compaction to all devices in Data-2016 and the results are shown under ``Endpoint compacted" in Table~\ref{table:firmwareResult}. 
Interestingly, this technique significantly enhances device identification; all state-4 devices transition to state-1. 
We observe that even with endpoint compaction, when MUD is unknown, the WeMo motion is incorrectly identified (as WeMo switch) at a high rate of $27.3$\%. 
This is expected; devices from the same manufacturer can get identified as one another when the endpoints are compacted.

In summary, if the identification process does not converge (or evolves very slowly) then our 
difference visualization and endpoint compaction allows a network operator to discover IoT devices running old firmware.

\noindent \textit{\textbf{Attacked or compromised device:}} We now evaluate the efficacy of our solution when IoT devices are under direct/reflection attacks or compromised by a botnet. 
We use traffic traces collected from our testbed in November 2017 (\ie Data-2017), comprising a number of volumetric attacks spanning reflection-and-amplification (\eg SNMP, SSDP, TCP SYN, Smurf), flooding (\eg TCP SYN, Fraggle, Ping of death), ARP spoofing, and port scanning.
The attacks were launched on four testbed IoT devices -- Belkin Netcam, WeMo motion, Samsung smart-cam and WeMo switch (listed in Table~\ref{table:attacsklist}). 

We initiated these attacks from the local network and from the Internet. 
For Internet-sourced attacks, port forwarding was enabled on the gateway (emulating malware behavior). 

We built a custom device type -- ``Senseme" \cite{wso2iots} --
using an Arduino Yun board communicating to the open-source WSO2 IoT cloud platform.
We built this device because our testbed IoT devices are all invulnerable to botnets.
This device has a temperature sensor and a bulb
and it periodically publishes the local temperature to its server and its bulb can be remotely controlled via the MQTT protocol \cite{senseme}. 
We generated the MUD profile of this device and then infected it with the Mirai botnet \cite{mirai}. 
We disabled the injection module of the Mirai code and only used its scanning module
to avoid harming others on the Internet.
A Mirai infected device scans random IP addresses on the Internet to find open telnet ports.

\begin{table}[t!]
	\centering
	\caption{List of attacks launched against our IoT devices \break(\textit{\textbf{L}: local, \textbf{D}: device, \textbf{I}: Internet}).}
	\label{table:attacsklist}
	\vspace{-0.2cm}
	\begin{adjustbox}{max width=0.44\textwidth}		
		\renewcommand{\arraystretch}{1.1}      					
		\begin{tabular}{|c|c||c|c|c|c||c|c|c|c|c|c|c|c|}
			\hline 
			\multicolumn{2}{|c||}{} & \multicolumn{4}{c||}{\textbf{Device}} & \multicolumn{6}{c|}{\textbf{Attack category}}\tabularnewline
			\hline 
			\multicolumn{2}{|c||}{Attacks} & \rotatebox{90}{WeMo motion} & \rotatebox{90}{WeMo switch} & \rotatebox{90}{Belkin cam} & \rotatebox{90}{Samsung cam} & \rotatebox{90}{L$\rightarrow$D} & \rotatebox{90}{L$\rightarrow$D$\rightarrow$L} & \rotatebox{90}{L$\rightarrow$D$\rightarrow$I} & \rotatebox{90}{I$\rightarrow$D$\rightarrow$I} & \rotatebox{90}{I$\rightarrow$D$\rightarrow$L} & \rotatebox{90}{I$\rightarrow$D}\tabularnewline
			\hline 
			\multirow{4}{*}{ \rotatebox{90}{Reflection}} & SNMP &  &  &  & \cmark &  & \cmark & \cmark & \cmark &  & \tabularnewline
			\cline{2-12} 
			& SSDP & \cmark &  & \cmark &  &  & \cmark & \cmark & \cmark &  & \tabularnewline
			\cline{2-12} 
			& TCP SYN & \cmark & \cmark & \cmark & \cmark &  & \cmark &  &  & \cmark & \tabularnewline
			\cline{2-12} 
			& Smurf & \cmark & \cmark & \cmark & \cmark &  & \cmark & \cmark &  &  & \tabularnewline
			\hline 
			\multirow{3}{*}{ \rotatebox{90}{Direct}} & TCP SYN & \cmark & \cmark & \cmark & \cmark & \cmark &  &  &  &  & \cmark\tabularnewline
			\cline{2-12} 
			& Fraggle & \cmark & \cmark & \cmark & \cmark & \cmark &  &  &  &  & \cmark\tabularnewline
			\cline{2-12} 
			& ICMP & \cmark & \cmark & \cmark & \cmark & \cmark &  &  &  &  & \tabularnewline
			\hline 
			\multicolumn{2}{|c||}{ARP spoof} & \cmark & \cmark & \cmark & \cmark & \cmark &  &  &  &  & \tabularnewline
			\hline 
			\multicolumn{2}{|c||}{Port Scan} & \cmark & \cmark & \cmark & \cmark & \cmark &  &  &  &  & \tabularnewline
			\hline 
		\end{tabular}
	\end{adjustbox}
\end{table}

\begin{figure}[t!]
	\centering
	\includegraphics[width=0.45\textwidth]{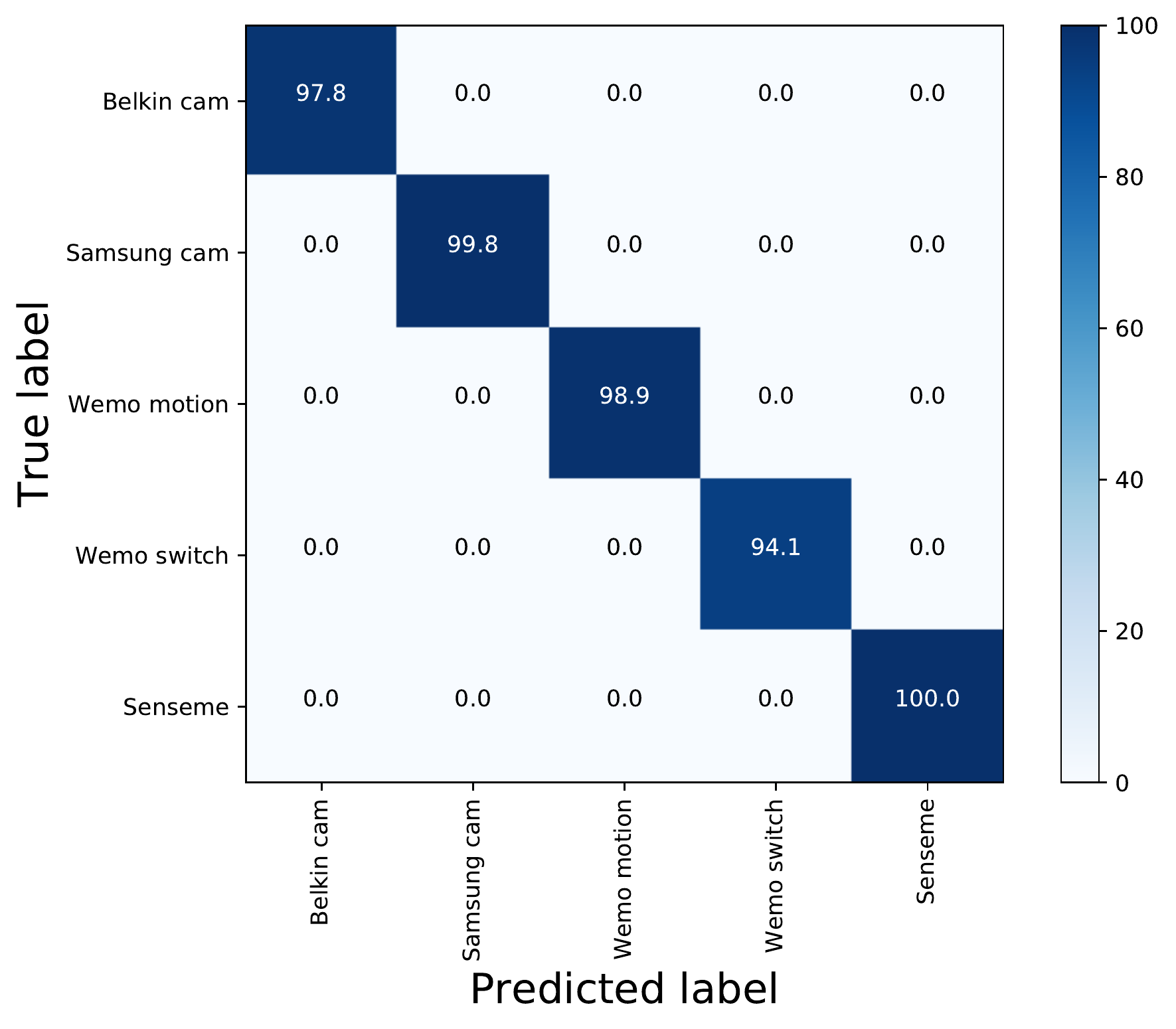}
	\vspace{-5mm}
	\caption{Partial confusion matrix for 5 devices only (testing with attack data 2017).}
	\label{fig:confMapAttack}
	\vspace{-4mm}
\end{figure}
We applied our threshold-based identification method to Data-2017 
and found that all devices were identified correctly with a high static similarity and low dynamic similarity (\ie high deviation). 
A partial confusion matrix for this is shown in Fig.~\ref{fig:confMapAttack}. 
The run-time profile of the Senseme quickly converges to the winner (with a high static similarity score)
because the device's MUD profile is fairly simple in terms of the branch count.
Other devices take longer to converge. 

\begin{figure}[t!]
	\centering
	\includegraphics[width=0.44\textwidth]{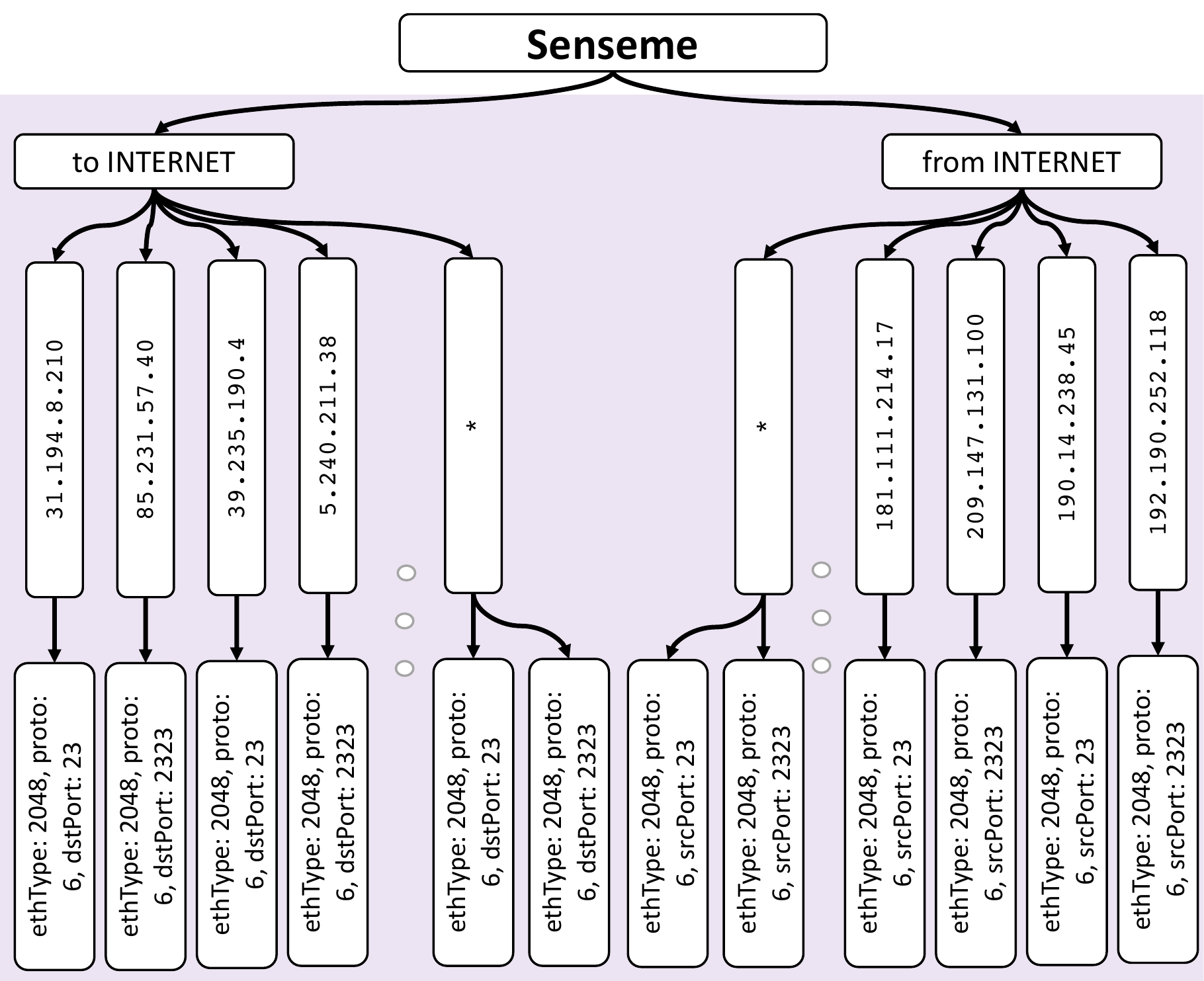}
	\vspace{-3mm}	
	\caption{Profile difference for the Mirai infected device.}
	\label{fig:miraiprofile}
	\vspace{-4mm}
\end{figure}

\begin{figure}[t!]
	\centering
	\includegraphics[width=0.35\textwidth]{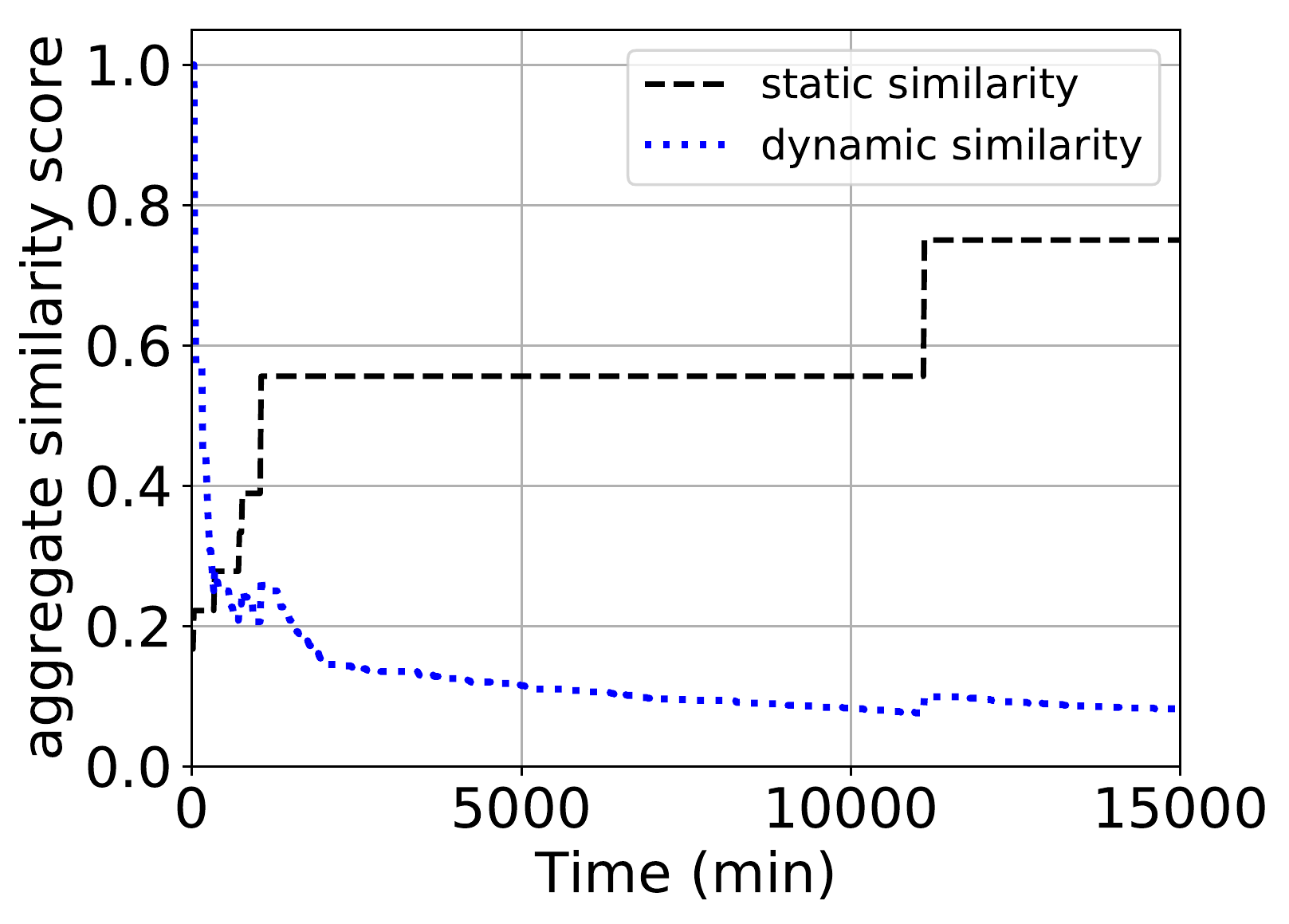}
	\vspace{-3mm}	
	\caption{Evolution of similarity scores for Belkin camera under attack.}
	\label{fig:belkinNetcamAttack}
	\vspace{-4mm}
\end{figure}

Various attacks have different impacts on the run-time profile of IoT devices. 
ARP spoofing and TCP SYN based attacks do not create new branches in a device profile's tree structure, hence, no deviation is captured. 
Fraggle, ICMP, Smurf, SSDP, and SNMP attacks result in only two additional flows, so a small deviation is captured. 
Port scans (botnet included) initiate a large deviation and cause an increasing number of endpoints to emerge in the tree structure at run-time. 
For example, the Mirai botnet scans 30 IP addresses per second, lowering the dynamic similarity to zero.
Fig.~\ref{fig:miraiprofile} shows the profile difference for the infected Senseme device at run-time.   
Lastly, we show in Fig.~\ref{fig:belkinNetcamAttack} the evolution of similarity scores for Belkin camera under  attack. It is seen that the static similarity slowly grows till it coverages to the correct winner -- according to Fig.~\ref{fig:confMapAttack} the first row, $2.2$\% of instances (only during the beginning of the process) did not converge to any winner. Instead, the dynamic similarity falls in time approaching to zero.

\vspace{-3mm}
\subsection{Performance of Monitoring Profiles}\label{sec:performMUDcheck}
We now quantify the performance of our scheme for real-time monitoring of IoT behavioral profiles. We use four metrics namely convergence time, memory usage, inspected packets, and number of flows.

\textbf{Convergence time:} Convergence time highly depends on type of the device and the similarity score thresholds. We note that the device network activity (\ie user interactions with the device) is an important factor for the convergence, since some IoTs (\eg Blipcare BP meter) do not communicate unless user interacts with the device. On the other hand, devices such as Awair air quality and WeMo motion sensor do not require any user interactions, and also cameras display a variety of communication patterns including device-to-device and device-to-Internet. 

Table~\ref{table:convergedTime} shows the convergence time (in minutes) for individual devices in our testbed, across the three datasets. For the Data-2018, all devices converge to their correct winner within a day -- the longest time taken to converge is 6 hours. This is primarily because that for this dataset we developed a script (using a touch replay tool running on a Samsung galaxy tablet connected to the same testbed) that automatically emulated the user interactions (via mobile app) with each of these IoT devices (\eg turning on/off the lightbulb, or checking the live view of the camera). Our script repeated every 6 hours.

\begin{table}[t!]
	\centering
	\caption{Convergence time (minutes) for all datasets.}\label{table:convergedTime}
	\vspace{-2mm}
	\begin{adjustbox}{max width=0.410\textwidth}
		\renewcommand{\arraystretch}{1.05}      			
		\begin{tabular}{|l|C{1.5cm}|C{1.5cm}|C{1.5cm}|}
			\hline 
			\textbf{Device} & \textbf{Data-2018} & \textbf{Data-2017} & \textbf{Data-2016}\tabularnewline
			\hline 
			\hline 
			Amazon Echo & 15 & - & 38355\tabularnewline
			\hline 
			August doorbell& 60 & - & \cellcolor{red!25}45\tabularnewline
			\hline 
			Awair air quality & 30 & - & 15\tabularnewline
			\hline 
			Belkin camera & 15 & 1065 & 105\tabularnewline
			\hline 
			Blipcare BPmeter & 15 & - & 15\tabularnewline
			\hline 
			Canary camera & 15 & - & 15\tabularnewline
			\hline 
			Chromecast & 15 & - & -\tabularnewline
			\hline 
			Dropcam & 360 & - & 15\tabularnewline
			\hline 
			Hello barbie & 15 & - & 15\tabularnewline
			\hline 
			HP printer & 105 & - & 15\tabularnewline
			\hline 
			Hue bulb & 15 & - &\cellcolor{red!25} 9315\tabularnewline
			\hline 
			iHome powerplug & 15 & - &\cellcolor{red!25} 165\tabularnewline
			\hline 
			LiFX bulb & 15 & - & 15\tabularnewline
			\hline 
			Nest smoke & 15 & - &\cellcolor{red!25} 15\tabularnewline
			\hline 
			Netatmo camera & 360 & - &\cellcolor{red!25} 1650\tabularnewline
			\hline 
			Netatmo weather& 15 & - & 15\tabularnewline
			\hline 
			Pixstar photoframe & 15 & - & 15\tabularnewline
			\hline 
			Ring doorbell & 30 & - & 45\tabularnewline
			\hline 
			Samsung smartcam & 15 & 525 & 15\tabularnewline
			\hline 
			Smart Things & 360 & - & 13785\tabularnewline
			\hline 
			TPlink camera & 30 & - & 15\tabularnewline
			\hline 
			TPlink plug & 30 & - & 15\tabularnewline
			\hline 
			Triby speaker & 15 & - & 330\tabularnewline
			\hline 
			WeMo motion & 15 & 360 & 15\tabularnewline
			\hline 
			WeMo switch & 15 & 2820 & 15\tabularnewline
			\hline 
			Withings cardio & 15 & - & -\tabularnewline
			\hline 
			Withings sleep & 15 & - & -\tabularnewline
			\hline 
			Senseme & - & 15 & 15\tabularnewline
			\hline 
		\end{tabular}
	\end{adjustbox}
	\vspace{-5mm}
\end{table}

Looking into Data-2017 column, it took up to 2 days to converge for WeMo switch as an example -- we only studied five devices under attack. The red cells under Data-2016 correspond to devices that converged due to endpoint compaction, similar to Fig.~\ref{fig:endpointCompaction}. Note that without compaction technique none of these devices (except Netatmo camera) converge to a winner -- Netatmo device required 4410 minutes to converge without compaction.
Similarly, it took a considerable amount of time for Smart Things, Hue bulb, and Amazon echo to converge -- when analyzed the data, we found that these three devices had no network activity (except a few flows during a short interval at the beginning of the data capture) till close to the convergence time when they became reasonably active on the network.


In real production networks, the operator may choose to begin with an upper bound threshold of convergence time for each device. If it does not converge within stipulated time, the phase of endpoint compaction can trigger with a corresponding time threshold which is a hard limit for the monitoring process. If the device identification is not concluded before the expected time, the operator may decide to quarantine it for further inspection.

\begin{table}[t!]
	\centering
	\caption{Performance metric calculated for Data-2018.}\label{table:performance}
	\vspace{-2mm}
	\begin{adjustbox}{max width=0.4450\textwidth}
		\renewcommand{\arraystretch}{1.05}      				
		\begin{tabular}{|l|C{1.7cm}|C{1.2cm}|C{1.2cm}|C{1.4cm}|}
			\hline 
			\textbf{Device}& \textbf{\# flows \break(per min)} & \textbf{\# packets (per min)} & \textbf{\# nodes (per min)} & \textbf{computing time (ms)}
			\tabularnewline
			\hline 
			\hline 
			Amazon Echo& 13.72 & 6.58  & 68.83 & 1.38\tabularnewline
			\hline 
			August doorbell& 20.11 & 13.44  & 65.84 & 1.71 \tabularnewline
			\hline 
			Awair air quality& 7.14 & 0.25  & 14.98 & 0.38\tabularnewline
			\hline 
			Belkin camera & 16.26 & 5.79  & 65.3 & 0.95\tabularnewline
			\hline 
			Blipcare BPmeter & 9.00  & 9.00 & 7.00 & 0.01\tabularnewline
			\hline 
			Canary camera & 13.51 & 3.27  & 25.51& 0.65\tabularnewline
			\hline 
			Chromecast  & 13.05 & 10.10  & 346.65 & 5.20\tabularnewline
			\hline 
			Dropcam & 7.02 & 0.04  & 17.87 & 0.45\tabularnewline
			\hline 
			Hello barbie & 5.86 & 3.72  & 9.52&0.72\tabularnewline
			\hline 
			HP printer & 5.05 & 2.12  & 38.63&0.74\tabularnewline
			\hline 
			Hue Bulb & 9.75 & 2.43  & 40.30&0.89\tabularnewline
			\hline 
			ihome powerplug & 6.87 & 0.79  & 16.99&0.49\tabularnewline
			\hline 
			lifx bulb & 8.65  & 1.60 & 18.86&0.50\tabularnewline
			\hline 
			Nest smoke  & 5.30 & 27.00  & 65.70 & 1.55\tabularnewline
			\hline 
			Netatmo camera & 8.35 & 0.98  & 67.96 & 1.20\tabularnewline
			\hline 
			Netatmo weather & 5.04 & 11.13  & 9.00 & 0.26\tabularnewline
			\hline 
			Pixstar photoframe & 5.05 & 2.62  & 16.88& 0.34\tabularnewline
			\hline 
			Ring doorbell & 5.02 & 2.39  & 25.94 & 0.43\tabularnewline
			\hline 
			Samsung smartcam & 10.37 & 1.34  & 209.98 & 2.00\tabularnewline
			\hline 
			Smart Things & 7.50 & 3.20  & 13.96 & 0.27 \tabularnewline
			\hline 
			TPlink camera  & 5.74 & 2.67 & 122.27 & 1.44 \tabularnewline
			\hline 
			TPlink plug & 5.07  & 3.96 & 26.49& 0.51\tabularnewline
			\hline 
			Triby speaker & 5.39 & 4.19  & 41.80& 0.75\tabularnewline
			\hline 
			WeMo motion  & 14.76 & 10.66 & 213.59& 2.97\tabularnewline
			\hline 
			WeMo switch & 6.54 & 4.46  & 225.99&5.20\tabularnewline
			\hline 
			Withings cardio & 5.57 & 11.00  & 9.00&0.15\tabularnewline
			\hline 
			Withings sleep & 21.00 & 27.00  & 22.00 &0.01\tabularnewline
			\hline 
		\end{tabular}
	\end{adjustbox}
	\vspace{-1mm}
\end{table}

\textbf{System performance:} For the performance of our system, we compute the following four metrics: average number of inspected packets, average number of flows, average number of nodes (in the device profile tree), and the computation time (for tree compaction, removing redundancy, and computing similarity scores) in our software tool. The average number of flows is an important metric for the operation of hardware switch with limited TCAM capacity, and other three metrics relate to the scalability of our software for run-time profile monitoring. Our results are shown in Table~\ref{table:performance}.

Starting from the first column, the average number of flows for each device is typically less than $10$, with the largest flow count of about $20$ for August doorbell. We note that this range of flow counts is easily manageable in an enterprise network setting with switches that are capable of handling millions of flow entries. In home networks, instead, where routers that can accommodate up to hundreds of flows, we may need to limit our monitoring process to a few devices (at a time), managing the TCAM constraint. 


Moving to the number of packets inspected, it is clearly seen that our approach is very effective by keeping the number of inspected packets at a minimum (\ie mostly less than $10$ packets per minute for each device). The computing time of our scheme solely depends on the number of nodes and the number of known MUD profiles. The time complexity of our method can be expressed as $\mathcal{O}(n.m.\log{}n)$ where $n$ is the number of branches in the profile tree and $m$ is the number MUD profiles we are checking against. We have reduced the time complexity for the search space by employing hashing and binary search tree techniques. For Chromescast as an example, the average computing time is $5.20$ ms, where we have on average $346$ nodes in its run-time profile. This can be further improved with use of parallelization whereby similarity scores are computed over individual branches. It is important to note that the computing time is upper-bounded since we set an upper bound limit on the count of tree branches generated at run-time.

Lastly, in terms of space, we need 40 Bytes of memory for each node of a tree. This means that for Chromecast, on average, less than 14 KB of memory is needed.  Additionally, all known MUD profiles are present in memory. Therefore, the space complexity heavily depends on the number of MUD profiles being checked.

\section{Conclusion}\label{sec:con}

In this paper, we have proposed a suite of tools that allows to automatically generate and formally verify IoT device behavior
and dynamically monitor IoT behavioral changes in a network. 
We demonstrate using these tools how the IETF MUD proposal can help reduce the effort needed to
dynamically identify and secure IoT devices.


\balance
\bibliographystyle{IEEEtran}
\bibliography{MUDcheck}

\begin{thebibliography}{10}
\providecommand{\url}[1]{#1}
\csname url@samestyle\endcsname
\providecommand{\newblock}{\relax}
\providecommand{\bibinfo}[2]{#2}
\providecommand{\BIBentrySTDinterwordspacing}{\spaceskip=0pt\relax}
\providecommand{\BIBentryALTinterwordstretchfactor}{4}
\providecommand{\BIBentryALTinterwordspacing}{\spaceskip=\fontdimen2\font plus
\BIBentryALTinterwordstretchfactor\fontdimen3\font minus
  \fontdimen4\font\relax}
\providecommand{\BIBforeignlanguage}[2]{{%
\expandafter\ifx\csname l@#1\endcsname\relax
\typeout{** WARNING: IEEEtran.bst: No hyphenation pattern has been}%
\typeout{** loaded for the language `#1'. Using the pattern for}%
\typeout{** the default language instead.}%
\else
\language=\csname l@#1\endcsname
\fi
#2}}
\providecommand{\BIBdecl}{\relax}
\BIBdecl

\bibitem{IoTSnP18}
A.~Hamza, D.~Ranathunga, H.~Habibi~Gharakheili, M.~Roughan, and V.~Sivaraman,
  ``{Clear As MUD: Generating, Validating and Applying IoT Behavioral
  Profiles},'' in \emph{Proc. ACM Workshop on IoT Security and Privacy (IoT
  S\&P)}, Budapest, Hungary, August 2018.

\bibitem{sachs2014}
G.~Sachs, ``{The Internet of Things: The Next Mega-Trend},'' [Online].
  Available: \url{www.goldmansachs.com/our-thinking/pages/internet-of-things/},
  2014.

\bibitem{Shodan}
J.~Matherly. (2018) {Shodan}. \url{https://www.shodan.io/}.

\bibitem{Dyn16}
S.~Hilton. (2016) {Dyn Analysis Summary Of Friday October 21 Attack}.
  \url{https://bit.ly/2xCr7WN}.

\bibitem{Wisec17}
M.~Lyu, D.~Sherratt, A.~Sivanathan, H.~Habibi~Gharakheili, A.~Radford, and
  V.~Sivaraman, ``{Quantifying the Reflective DDoS Attack Capability of
  Household IoT Devices},'' in \emph{Proc. ACM WiSec}, Boston, Massachusetts,
  July 2017.

\bibitem{DHS16}
U.~D. of~Homeland~Security. (2016) {Strategic Principles For Securing the
  Internet of Things (IoT)}. \url{https://bit.ly/2eXOGzV}.

\bibitem{NIST16}
NIST. (2016) {Systems Security Engineering}. \url{https://bit.ly/2tak6fP}.

\bibitem{ENISA17}
E.~U. A.~F. Network and I.~Security. (2017) {Communication network dependencies
  for ICS/SCADA Systems}.
  \url{https://www.enisa.europa.eu/publications/ics-scada-dependencies}.

\bibitem{FCC16}
FCC. (2016) {Federal Communications Comssion Response 12-05-2016}.
  \url{https://bit.ly/2gUztSv}.

\bibitem{ietfMUD18}
\BIBentryALTinterwordspacing
E.~Lear, R.~Droms, and D.~Romascanu, ``Manufacturer usage description
  specification (work in progress),'' Working Draft, IETF Secretariat,
  Internet-Draft draft-ietf-opsawg-mud-18, January 2018. [Online]. Available:
  \url{http://www.ietf.org/internet-drafts/draft-ietf-opsawg-mud-18.txt}
\BIBentrySTDinterwordspacing

\bibitem{hamza2018}
A.~Hamza, D.~Ranathunga, H.~Habibi~Gharakheili, M.~Roughan, and V.~Sivaraman,
  ``{Clear as MUD: Generating, Validating and Applying IoT Behavioral
  Profiles},'' in \emph{Proc. ACM Workshop on IoT Security and Privacy (IoT
  S\&P)}, Budapest, Hungary, Aug 2018.

\bibitem{mudgenerator}
A.~Hamza. (2018) {MUDgee}. \url{https://github.com/ayyoob/mudgee}.

\bibitem{Survey17}
D.~M. Mendez, I.~Papapanagiotou, and B.~Yang, ``{Internet of Things: Survey on
  Security and Privacy},'' \emph{CoRR}, vol. abs/1707.01879, 2017.

\bibitem{IoTSnp17}
F.~Loi, A.~Sivanathan, H.~H. Gharakheili, A.~Radford, and V.~Sivaraman,
  ``{Systematically Evaluating Security and Privacy for Consumer IoT
  Devices},'' in \emph{Proc. ACM IoT S\&P}, Dallas, Texas, USA, Nov 2017.

\bibitem{CiscoReport18}
C.~Systems, ``{Cisco 2018 Annual Cybersecurity Report},'' Tech. Rep., 2018.

\bibitem{f5Labs17}
S.~Boddy and J.~Shattuck, ``{The Hunt for IoT: The Rise of Thingbots},'' {F5
  Labs}, Tech. Rep., July 2017.

\bibitem{sivaraman2016smart}
V.~Sivaraman, D.~Chan, D.~Earl, and R.~Boreli, ``{Smart-Phones Attacking
  Smart-Homes},'' in \emph{Proc. ACM WiSec}, Darmstadt, Germany, July 2016.

\bibitem{SonyCam}
P.~World. (2018) {Backdoor accounts found in 80 Sony IP security camera
  models}. \url{https://bit.ly/2GbKejk}.

\bibitem{Insecam18}
(2018) {MUD maker}. \url{http://www.insecam.org/en/bycountry/US/}.

\bibitem{SmartCity17}
A.~Sivanathan, D.~Sherratt, H.~Habibi~Gharakheili, A.~Radford, C.~Wijenayake,
  A.~Vishwanath, and V.~Sivaraman, ``{Characterizing and classifying IoT
  traffic in smart cities and campuses},'' in \emph{Proc. IEEE INFOCOM workshop
  on SmartCity}, Atlanta, Georgia, USA, May 2017.

\bibitem{wool2010}
A.~Wool, ``{Trends in Firewall Configuration Errors: Measuring the Holes in
  Swiss Cheese},'' \emph{IEEE Internet Computing}, vol.~14, no.~4, pp. 58--65,
  2010.

\bibitem{ranathunga2016T}
D.~Ranathunga, M.~Roughan, H.~Nguyen, P.~Kernick, and N.~Falkner, ``Case
  studies of scada firewall configurations and the implications for best
  practices,'' \emph{IEEE Transactions on Network and Service Management},
  vol.~13, pp. 871--884, 2016.

\bibitem{ranathunga2016G}
D.~Ranathunga, M.~Roughan, P.~Kernick, N.~Falkner, H.~Nguyen, M.~Mihailescu,
  and M.~McClintock, ``{Verifiable Policy-defined Networking for Security
  Management},'' in \emph{SECRYPT}, 2016, pp. 344--351.

\bibitem{basu2007}
A.~Basu and R.~Blanning, \emph{Metagraphs and their applications}.\hskip 1em
  plus 0.5em minus 0.4em\relax Springer Science \& Business Media, 2007,
  vol.~15.

\bibitem{meidan2017}
Y.~Meidan, M.~Bohadana, A.~Shabtai, M.~Ochoa, N.~O. Tippenhauer, J.~D.
  Guarnizo, and Y.~Elovici, ``Detection of unauthorized iot devices using
  machine learning techniques,'' \emph{arXiv preprint arXiv:1709.04647}, 2017.

\bibitem{sivanathan2018}
A.~Sivanathan, H.~Habibi~Gharakheili, F.~Loi, A.~Radford, C.~Wijenayake,
  A.~Vishwanath, and V.~Sivaraman, ``{Classifying IoT Devices in Smart
  Environments Using Network Traffic Characteristics},'' \emph{IEEE
  Transactions on Mobile Computing}, 2018.

\bibitem{wool2004}
A.~Wool, ``A quantitative study of firewall configuration errors,'' \emph{IEEE
  Computer}, vol.~37, no.~6, pp. 62--67, 2004.

\bibitem{al2005}
E.~Al-Shaer, H.~Hamed, R.~Boutaba, and M.~Hasan, ``Conflict classification and
  analysis of distributed firewall policies,'' \emph{IEEE JSAC}, vol.~23,
  no.~10, pp. 2069--2084, 2005.

\bibitem{cisco2013}
{Cisco Systems}, \emph{Cisco {ASA} Series {CLI} Configuration Guide, 9.0},
  Cisco Systems, Inc., 2013.

\bibitem{juniper2016}
{Juniper Networks, Inc.}, \emph{Getting Started Guide for the Branch {SRX}
  Series}, 1133 Innovation Way, Sunnyvale, CA 94089, USA, 2016.

\bibitem{palo2017}
{Palo Alto Networks, Inc.}, \emph{PAN-OS Administrator's Guide, 8.0}, 4401
  Great America Parkway, Santa Clara, CA 95054, USA, 2017.

\bibitem{ranathunga2017}
D.~Ranathunga, H.~Nguyen, and M.~Roughan, ``Mgtoolkit: A python package for
  implementing metagraphs,'' \emph{SoftwareX}, vol.~6, pp. 91--93, 2017.

\bibitem{ranathunga2016P}
D.~Ranathunga, M.~Roughan, P.~Kernick, and N.~Falkner, ``{Malachite: Firewall
  policy comparison},'' in \emph{IEEE Symposium on Computers and Communication
  (ISCC)}, June 2016, pp. 310--317.

\bibitem{stouffer2008}
K.~Stouffer, J.~Falco, and K.~Scarfone, ``Guide to {I}ndustrial {C}ontrol
  {S}ystems ({I}{C}{S}) security,'' \emph{NIST Special Publication}, vol. 800,
  no.~82, pp. 16--16, 2008.

\bibitem{byres2005}
E.~Byres, J.~Karsch, and J.~Carter, ``{NISCC} good practice guide on firewall
  deployment for {SCADA} and process control networks,'' \emph{NISCC}, 2005.

\bibitem{plonka2003}
D.~Plonka. (2013) {Flawed Routers Flood University of Wisconsin Internet Time
  Server}. \url{www.pages.cs.wisc.edu/~plonka/netgear-sntp/}.

\bibitem{sdnsimulator}
\BIBentryALTinterwordspacing
A.~Hamza. (2018) {SDN pcap simulator}. [Online]. Available:
  \url{https://github.com/ayyoob/sdn-pcap-simulator}
\BIBentrySTDinterwordspacing

\bibitem{MUDids2018}
A.~Hamza, H.~Habibi~Gharakheili, and V.~Sivaraman, ``Combining mud policies
  with sdn for iot intrusion detection,'' in \emph{Proc. ACM Workshop on IoT
  Security and Privacy (IoT S\&P)}, Budapest, Hungary, August 2018.

\bibitem{wso2iots}
\BIBentryALTinterwordspacing
(2018) {WSO2 IoT Server}. [Online]. Available: \url{https://wso2.com/iot}
\BIBentrySTDinterwordspacing

\bibitem{senseme}
\BIBentryALTinterwordspacing
(2018) {SenseMe}. [Online]. Available:
  \url{https://github.com/wso2/samples-iots/tree/master/SenseMe}
\BIBentrySTDinterwordspacing

\bibitem{mirai}
\BIBentryALTinterwordspacing
(2018) {Mirai botnet}. [Online]. Available:
  \url{https://github.com/jgamblin/Mirai-Source-Code}
\BIBentrySTDinterwordspacing

\end{thebibliography}


\addcontentsline{toc}{section}{List of Acronyms}
\begin{acronym}[TDMA]
        \setlength{\itemsep}{-\parsep}
        \acro{ACLs}{Access Control Lists}
        \acro{STUN}{Session Traversal Utilities for NAT}
\end{acronym}

\vspace{-1cm}

\end{document}